\documentclass{rspublic}

\usepackage{amsmath}
\usepackage{amssymb,amsfonts}
\usepackage{bm}
\usepackage[mathcal]{euscript}
\usepackage{graphicx}
\graphicspath{{figs/}{figs_lo/}}
\usepackage{subfigure}
\usepackage{floatflt}
\usepackage{natbib}

\bibpunct{(}{)}{;}{a}{}{}

\newcommand{\commentjlt}[1]{}
\newcommand{\comment}[1]{}

\newcommand{\mathnotation}[2]{\newcommand{#1}{\ensuremath{#2}}}

\mathnotation{\ldef}{\mathrel{\raisebox{.069ex}{:}\!\!=}}
\mathnotation{\rdef}{\mathrel{=\!\!\raisebox{.069ex}{:}}}
\mathnotation{\nn}{n}				
\mathnotation{\R}{R}				
\mathnotation{\f}{f}				
\mathnotation{\Br}{B}				
\mathnotation{\nword}{k}

\begin{document}

\title[Topology, Braids, and Mixing in Fluids]
      {Topology, Braids, and Mixing in Fluids}

\author[J.-L. Thiffeault and M. D. Finn]
       {Jean-Luc Thiffeault and Matthew D. Finn}

\affiliation{Department of Mathematics,
  Imperial College London, London SW7 2AZ, UK}

\label{firstpage}

\maketitle

\begin{abstract}{chaotic mixing, topological chaos}
Stirring of fluid with moving rods is necessary in many practical applications
to achieve homogeneity.  These rods are topological obstacles that force
stretching of fluid elements.  The resulting stretching and folding is
commonly observed as filaments and striations, and is a precursor to mixing.
In a space-time diagram, the trajectories of the rods form a braid, and the
properties of this braid impose a minimal complexity in the flow.  We review
the topological viewpoint of fluid mixing, and discuss how braids can be used
to diagnose mixing and construct efficient mixing devices.  We introduce a
new, realisable design for a mixing device, the \emph{silver mixer}, based on
these principles.
\end{abstract}

\section{Introduction}

\begin{figure}
\centering{\includegraphics[width=.45\textwidth]{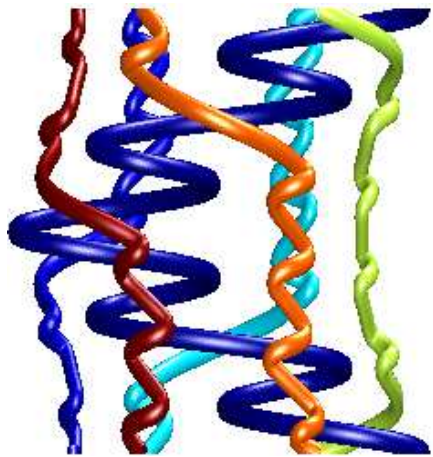}}
\caption{The braid formed by the world lines of a rod and several periodic
  islands in a mixing device.}
\label{fig:tubeplot}
\end{figure}

In recent years, starting with the work of~\citet{Boyland2000}, methods from
topology and braid theory have been applied to the analysis of mixing in flows
with great success.  Consider a mixing device consisting of rods moving in a
fluid.  If the horizontal position of the rods is plotted in a
three-dimensional graph, with time the vertical axis, one obtains a `spaghetti
plot' of the world lines of the rods, similar to figure~\ref{fig:tubeplot}.
Because the rods are material objects, the world lines cannot intersect each
other.  The resulting graph thus describes a braid, in the mathematical sense
of a bundle of strands that are not allowed to cross each other.
%
%

The analysis of this braid yields important information about the mixing
properties of the flow.  For instance, it can be shown that if the braid
possesses a positive \emph{topological entropy}, then there must exist a
region in the surrounding fluid that exhibits chaotic trajectories.  The
presence of chaos is a consequence of the topology of the rod motion, and is
guaranteed no matter which dynamical equations the fluid obeys, or whether the
flow is laminar or turbulent---hence the name \emph{topological chaos}.  It is
well-known~\citep{Aref1984} that chaotic trajectories are very good for
mixing, especially in very viscous flows where turbulence is nonexistent.

The reasoning for the presence of chaos involves material lines in the fluid,
which are lines that are assumed to move with the fluid, in the manner that a
nondiffusing blob of dye would.  A material line is dragged and stretched by
the fluid motion, but it cannot cross the physical rods.  Hence, the material
line inherits the complexity of the rod motion because it `snags' on the rods.

But physical rods are not necessary for topological chaos.  In two dimensions,
any particle orbit is an obstacle to material lines~\citep{Thiffeault2005}.
Periodic orbits have been particularly useful in characterising topological
chaos. Perhaps paradoxically, the most appealing periodic orbits for this
purpose are \emph{regular islands}, also known as elliptic regions.
Figure~\ref{fig:tubeplot} shows the motion of a single rod as well as
regular islands created in a viscous flow by the motion of the rod.
Topologically speaking, there is no difference between the rod and the
islands, and the braid shown here has positive topological
entropy~\citep{Gouillart2006}.  This opens up the intriguing possibility of
optimising mixing devices by altering the braid formed by the motion of rods
and islands.

In this article we propose to (i) Briefly review the important issues for
mixing in fluids (\S\ref{sec:snm}); (ii) Describe the theory behind braids and
topological chaos (\S\ref{sec:topofluid}--\S\ref{sec:diag}); and (iii) Apply
braid optimisation techniques to real mixing devices (\S\ref{sec:opt}).  Along
the way we will mention open problems, and in \S\ref{sec:future} we will
discuss future avenues for exploration, and where the challenges lie.

\section{Stirring and Mixing in Fluids}
\label{sec:snm}

It comes as a surprise to many that mixing is actually a proper field of
study.  After all, how much of a mathematical challenge can stirring milk in a
teacup present?  Well, quite a difficult one, actually!  For the particular
case of the teacup, \emph{stirring} creates turbulence, and turbulent flows
are usually extremely good at \emph{mixing}.  Turbulence is hard---if not
impossible---to understand, so we are already in dangerous territory.

But it gets worse: the teacup is a poor example because there is not much to
achieving good mixing: a flick of the wrist will usually suffice.  But there
are many other situations of practical interest where this is not the case,
for various reasons.  The basic setting is the same: given some quantity
(e.g., milk, temperature, moisture, salt, dye\dots, usually referred to as the
\emph{scalar field}) that is \emph{transported} by a fluid (e.g., air or
water), how does the concentration of that substance evolve in time?  But from
there very different questions can arise:
\begin{itemize}
\item Does the scalar concentration tend to a constant distribution, and if
  so, how rapidly?
\item Does the scalar eventually fill the entire domain, or are
  there \emph{transport barriers} that prevent this?
\item How much energy is required to stir the fluid?
\end{itemize}
For the teacup, the answer to these questions are: yes, fast; yes; and not
very much at all.  Note that stirring is the mechanical process that moves the
fluid, and mixing is the tendency of the scalar field to homogenise.  In this
paper we shall use the two words interchangeably.

Let us consider two extreme cases that are more challenging than the teacup.
First, oceanic flows: the ocean is preferentially heated at the equator; this
causes evaporation, which leads to higher salt concentrations.  But clearly
the ocean is not becoming hotter and saltier at the equator, so there must be
a mechanism that redistributes these two scalars over the globe.  One
candidate is molecular diffusion, which all scalars undergo, but that is
utterly negligible---it would take longer than the age of the earth to
redistribute the heat and salt in this way.  The primary mode of
redistribution is by far transport by oceanic currents.  In this case the
scalars are \emph{active} rather than \emph{passive}, since they influence the
flow itself because of the different weight of warm, salty water compared to
cold, fresh water.  But for climate modelling it is crucial to know how fast
the global redistribution of heat and salt occurs.  The number of effects that
play a role is astounding: tides, storms, the Earth's rotation, coupling to
the atmosphere, bottom topography, etc.

At the opposite end of the spectrum, consider \emph{lab-on-a-chip}
applications.  In this case, the fluid motion takes place at micrometre scales
in grooves etched on the surface of microchips.  At these scales, the motion
of a fluid like water behaves as a viscous fluid: turbulence is impractical to
achieve.  In these applications one wants to take a standard laboratory
manipulation and reduce it in size so it can be embedded in a small, hand-held
device for, say, forensic work.  But many applications require mixing: for
instance, a DNA sample needs to be mixed with the appropriate reagent.  The
problem is that the fluid motion is so regular that mixing is very difficult,
and the molecular diffusion of DNA is very slow (larger molecules diffuse more
slowly, and DNA is very large indeed by molecular standards).  This is where
\emph{chaotic mixing} becomes the best option, and the field has undergone a
renaissance because of lab-on-a-chip
applications~\citep{Stone2001,Whitesides2001}.

It was~\citet{Henon1966} who first realised that steady three-dimensional
flows could have chaotic trajectories.  This is a counterintuitive result: the
flow pattern is not changing in time, but if one starts two particle
trajectories close to each other they diverge exponentially, at a rate given
by the so-called \emph{Lyapunov exponent} of the flow.  \citet{Aref1984}
realised the same thing for two-dimensional flows with time dependence, and
also saw the advantage of this for fluid mixing, coining the term
\emph{chaotic advection}.  If the flow is chaotic, it means that fluid
particles rapidly become uncorrelated and forget about each other's
whereabouts.  But that is exactly what it means for a scalar to be mixed: the
initial concentration field is forgotten.%
\footnote{We are leaving out the crucial role of molecular diffusion in
  ultimately achieving this homogenisation.}
In essence, chaotic advection (also called chaotic mixing when diffusion is
implicitly assumed to act) can potentially achieve the same result as
turbulence as far as mixing is concerned, but with much simpler fluid motion
and at a lower energy cost.

A quantity that is closely related to the Lyapunov exponent is the
\emph{line-stretching exponent}: it gives the asymptotic growth rate of
material lines.  It is always greater than or equal to the Lyapunov exponent,
and the mismatch between the two is a measure of the nonuniformity of
stretching in the flow.  In two-dimensional smooth flows the line-stretching
exponent, maximised over all possible initial material lines, is equal to the
topological entropy of the flow \citep{Bowen1978,Franks1988,Newhouse1993}.  In
this paper we will use the topological entropy as a measure of mixing
quality, but many other measures are available~\citep{MattFinn2004}.

Here we will be concerned with time-dependent two-dimensional fluid motions,
as in \citet{Aref1984}.  The main application we have in mind is to batch
stirring devices, where very viscous fluid in a vat is stirred by rods.  The
motion of such fluid is approximately two-dimensional.  We will see that the
choice of motion of the rods (and not the speed) is crucial to achieving good
mixing.

\section{A Topological View of Fluid Motion}
\label{sec:topofluid}

\subsection{Isotopy Classes}

Having covered the physical background of mixing in \S\ref{sec:snm}, we now
turn to a mathematical description of fluid motion.  This section follows
closely \citet{Boyland2000}.  We consider the periodic motion of a
two-dimensional fluid in some domain~$\R_\nn$ with~$\nn$ identical stirring
rods, as in figure~\ref{fig:rods_s1s-2s3s-2} for four rods.  The rods undergo
some prescribed motion in one cycle and return to their initial configuration,
possibly having been permuted.

Fluid elements in $\R_\nn$ are dragged along by the rods, and their motion
after one period is given by a map $\f:\R_\nn \rightarrow R_\nn$.  We assume
this map is a diffeomorphism (smooth with smooth inverse), as is typical of
physical situations.  In practice, this diffeomorphism is obtained by solving
a set of fluid equations, such as the Stokes equations for a viscous fluid or
the Navier--Stokes equations when inertia is important.  But here we take $\f$
to be arbitrary, so that it could represent motions that are far from physical
and do not satisfy any particular fluid equations.  Rather, we will seek to
classify~$\f$ according to its topological properties, regardless of the type
of fluid it arises from.

The topological properties of $\f$ are encoded in its \emph{isotopy class}.
Two maps are \emph{isotopic} if they can be continuously deformed into each
other, allowing possibly for rotation of the boundaries.  If the fluid is
ideal, then rotation of the boundaries is irrelevant.  If a map $\f$ arises
from motions that only involve rotation of the rods or the external boundary,
then $\f$ is said to be \emph{isotopic to the identity}.  For instance, the
journal bearing flow~\citep{Aref1986,Chaiken1986}, consisting of two rotating
off-centre cylinders, leads to chaotic fluid motions that are isotopic to the
identity ($\nn=1$ in this case).  The set of all maps isotopic to $\f$ form
its isotopy class.

If $\nn=0$, $1$, or $2$, then all fluid motions are isotopic to the identity,
no matter how the rods are moved.  The key insight in \citet{Boyland2000} is
that for~$\nn\ge3$ there are fluid motions not isotopic to the identity.
These, we will see, lead to chaotic dynamics that are complex in a profound
topological sense--\emph{topological chaos}.

\subsection{The Thurston--Nielsen Classification}
\label{sec:TN}

Given a diffeomorphism $\f$, what possible isotopy classes can it belong to?
For two-dimensional compact manifolds, the \emph{Thurston--Nielsen
classification theorem} \citep{Thurston1988,Fathi1979} tells us that $\f$ is
isotopic to $\f'$, the \emph{Thurston--Nielsen representative}, where $\f'$ is
one of three types of mapping:
\begin{enumerate}
\item \emph{finite-order}: if~$\f'$ is repeated enough times, the resulting
  diffeomorphism is the identity;
\item \emph{pseudo-Anosov} (pA): $\f'$ stretches the fluid elements by a
  factor~$\lambda > 1$, so that repeated application gives exponential
  stretching; $\lambda$ is called the \emph{dilatation} of~$\f'$,
  and~$\log\lambda$ is its topological entropy.
\item \emph{reducible}: $\f'$ leaves a family of curves invariant, and these
  curves delimit subregions that are of type 1 or 2.
\end{enumerate}
Note that we have omitted some technical details---see \citet{Boyland1994} for
a full exposition.  For our purposes, the second of these classes (pA) is most
important.  Anosov diffeomorphisms are the prototypical chaotic maps: they
stretch uniformly everywhere.  The most famous example is Arnold's cat map on
the torus~\citep{ArnoldAvez}.  A pseudo-Anosov map allows for a finite number
of singularities in the stable and unstable foliations of the map.

Our goal in designing an efficient mixer is now clear: we want to induce a
diffeomorphism $\f$ that is either isotopic to a pseudo-Anosov map, or splits
$\R_\nn$ into subregions that include type 2 components.  The \emph{isotopy
stability theorem} of Handel~\citep{Handel1985} guarantees that $\f$ has
dynamics that are at least as complicated as the pA map in its isotopy class.
In the next section we will see that specifying isotopy classes is
conveniently achieved with \emph{braids}.

\section{Enforcing Chaos with Braids}
\label{sec:enforcing}

\subsection{Braids Label Isotopy Classes}

\begin{figure}
\begin{center}
\includegraphics[width=.65\textwidth]{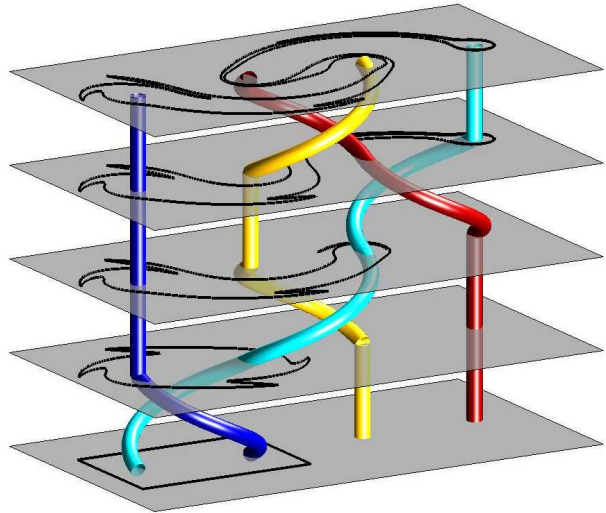}
\end{center}
\caption{The trajectories of four stirrers define a braid on~$\nn=4$ strands
in a space-time diagram, with time flowing from bottom to top.  In terms of
generators, the braid is
written~$\sigma_1\sigma_{2}^{-1}\sigma_3\sigma_{2}^{-1}$, read from left to
right. First (bottom) the leftmost two rods are interchanged clockwise
($\sigma_{1}$), then the middle two rods are interchanged counterclockwise
($\sigma_{2}^{-1}$), and so on.  The evolution of a material line (black
rectangle initially) advected by the flow is also shown.}
\label{fig:braidpicture}
\end{figure}

In \S\ref{sec:topofluid} we saw that the isotopy class of $\f$, a
diffeomorphism representing periodic fluid motion, is one of three types.  Now
we introduce a convenient way to label these isotopy classes---braids.  To see
intuitively how braids come in, lift the stirrer motions to a
three-dimensional space-time plot (the `world lines' of the rods), as in
figure~\ref{fig:braidpicture}.
The resulting plot is called a \emph{physical braid}.  Two braids are
equivalent if they can be deformed into each other with no strands
crossing other strands or boundaries.

We can pass from physical braids to an algebraic description of braids by
introducing \emph{generators}.  Assume without loss of generality that all the
stirrers lie on a line.  The generator $\sigma_i$, $i=1,\ldots,\nn-1$, denotes
the clockwise interchange of the $i$th rod with the $(i+1)$th rod along a
circular path, where~$i$ is the position of a rod counting from left to right.
The counterclockwise interchange is denoted~$\sigma_i^{-1}$.  (See
figure~\ref{fig:braidpicture}.)  We can write consecutive interchanges
as~$\sigma_1\sigma_{2}^{-1}$, where the generators are read from left to right
(this convention differs from \citet{Boyland2000}).  This composition law
allows us to define the \emph{braid group on $\nn$ strands}, $\Br_\nn$, with
the identity given by disentangled strands.  In order that braid group
elements correspond to physical braids, they must satisfy the additional
relations~$\sigma_i\sigma_{i+1}\sigma_i = \sigma_{i+1}\sigma_i\sigma_{i+1}$
and $\sigma_i\sigma_j = \sigma_j\sigma_i$, for $|i-j|>1$~\citep{Birman1975}.

The crucial role of braids as an organising tool is that they correspond to
isotopy classes in~$\R_\nn$.  (There is a subtlety involving possible
rotations of the outer boundary, so that one usually speaks of \emph{braid
types} specifying isotopy classes rather than braids~\citep{Boyland2000}.)  We
assign a braid to stirrer motions by the diagrammatic approach employed in
figure~\ref{fig:braidpicture}, and to every braid we can assign stirrer
motions by reversing the process.

\subsection{Good Braids and Bad Braids}
\label{sec:goodbad}

So what is to be learned from the braid description of stirrer motions?  We
can tell directly from the braid which isotopy class the motion belongs to,
using the train-track algorithm of~\citet{Bestvina1995}.  In this paper, we
have used a very convenient implementation of this algorithm written by Toby
Hall,%
\footnote{See
\texttt{http://www.liv.ac.uk/maths/PURE/MIN\_SET/CONTENT/members/%
T\_Hall.html}.}
in which one simply types in the generators, and the program returns the
isotopy class and dilatation.  We will return to the advantages and
disadvantages of the Bestvina--Handel algorithm in
\S\ref{sec:diag}\ref{sec:chaorb}.

\begin{figure}
\begin{center}
\parbox[t]{3.0cm}{\includegraphics[width=2.85cm]{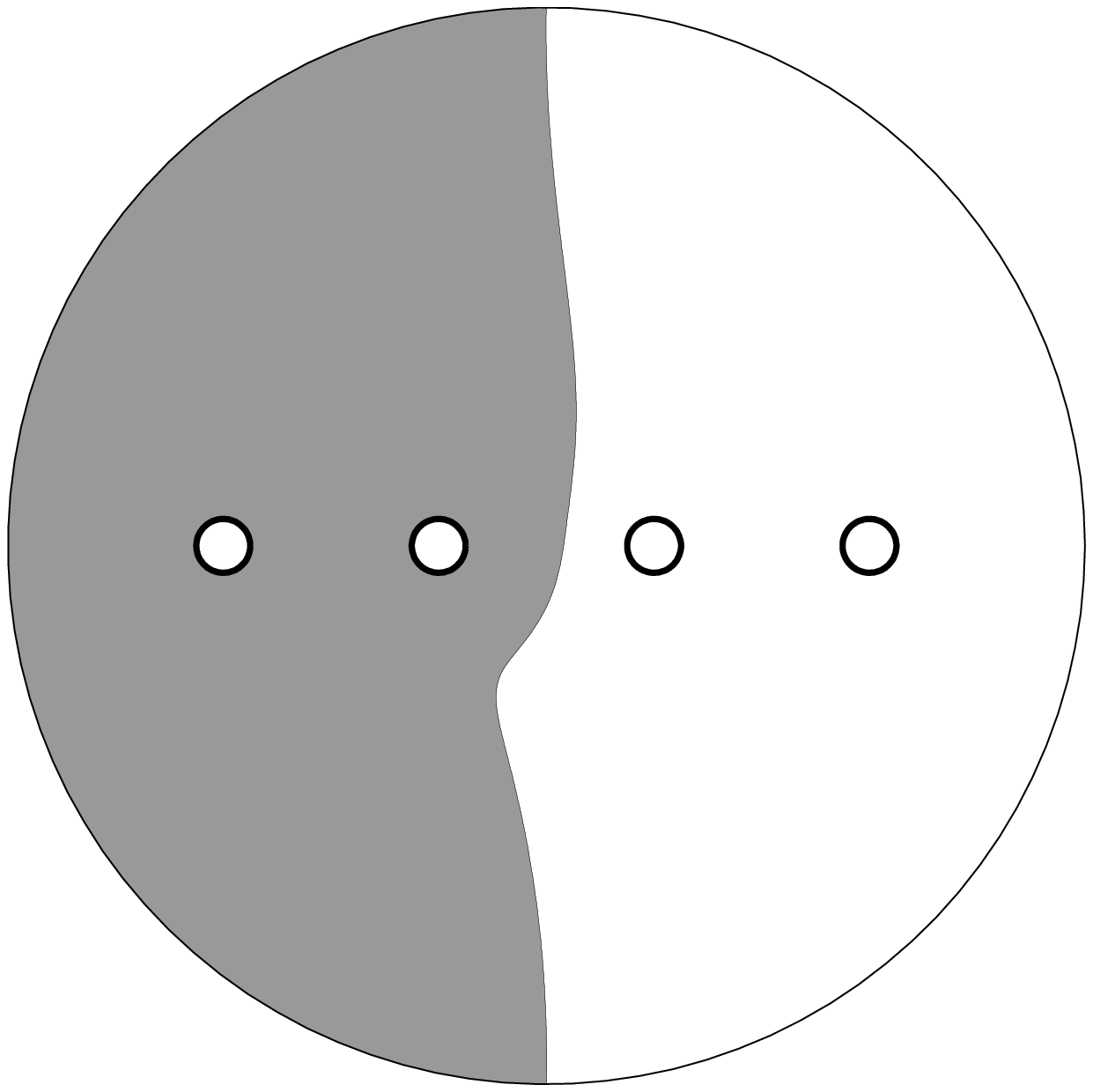}}
\parbox[t]{3.0cm}{\includegraphics[width=2.85cm]{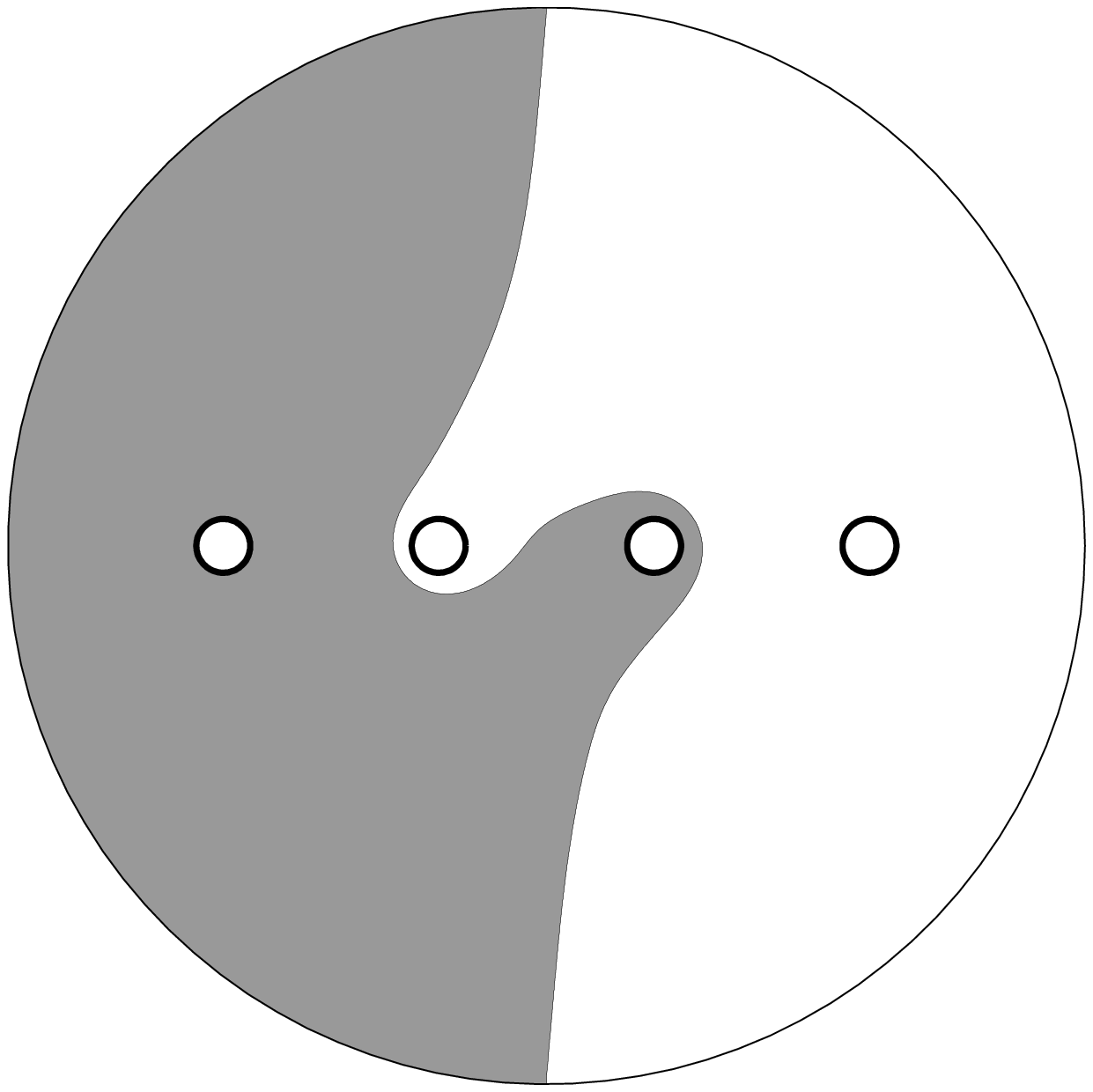}}
\parbox[t]{3.0cm}{\includegraphics[width=2.85cm]{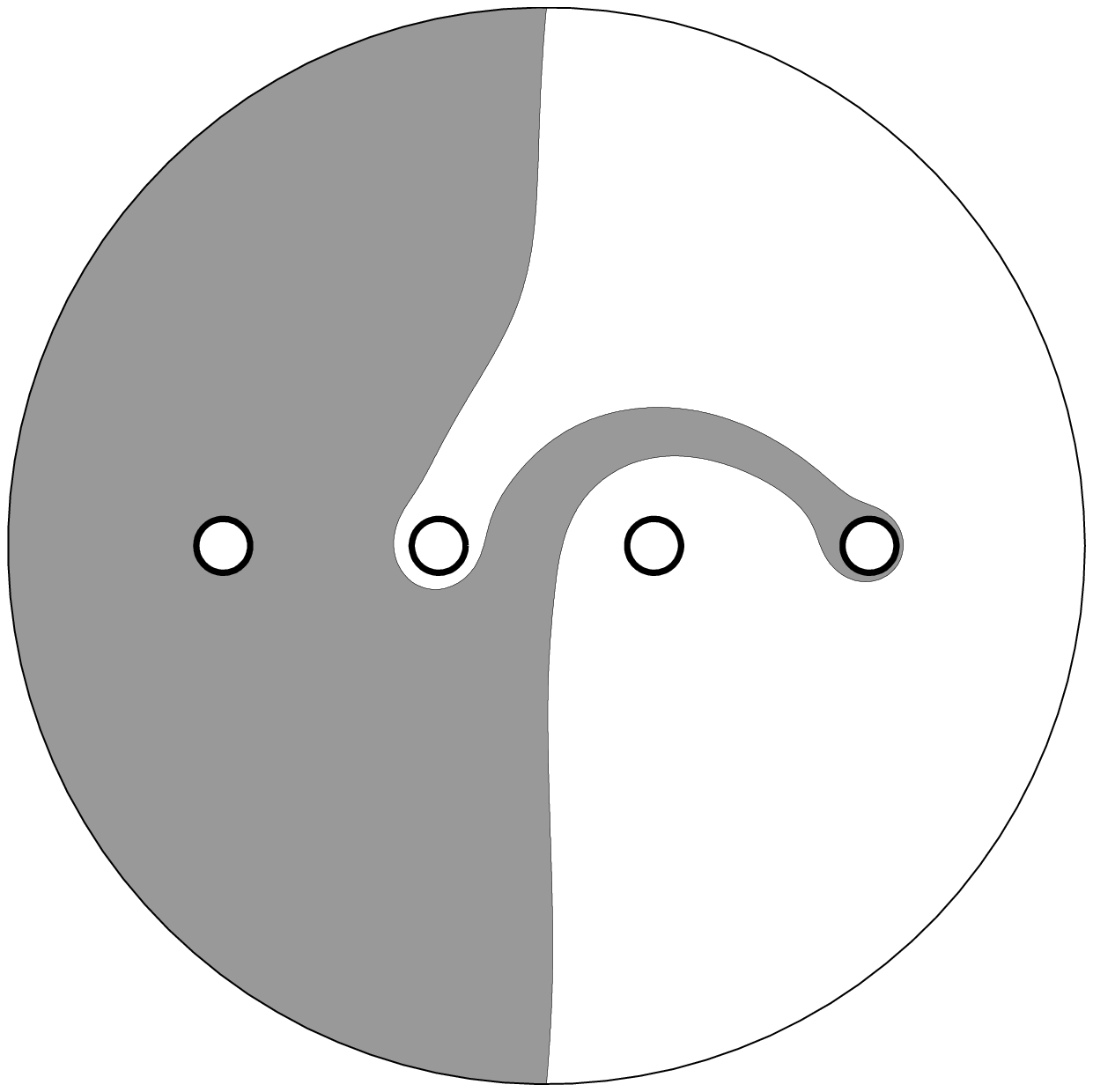}}
\parbox[t]{3.0cm}{\includegraphics[width=2.85cm]{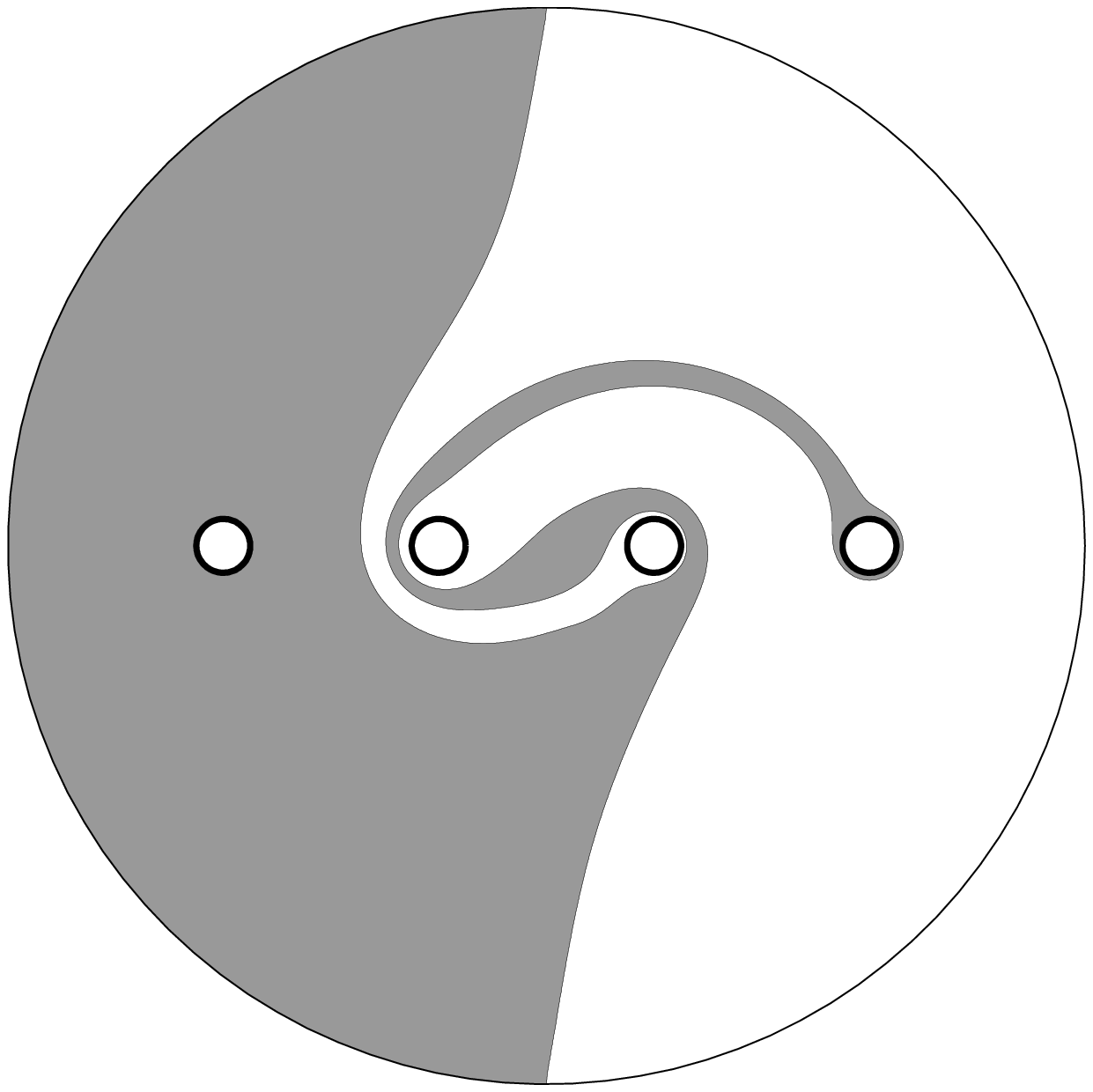}}

\smallskip

\parbox[t]{3.0cm}{\includegraphics[width=2.85cm]{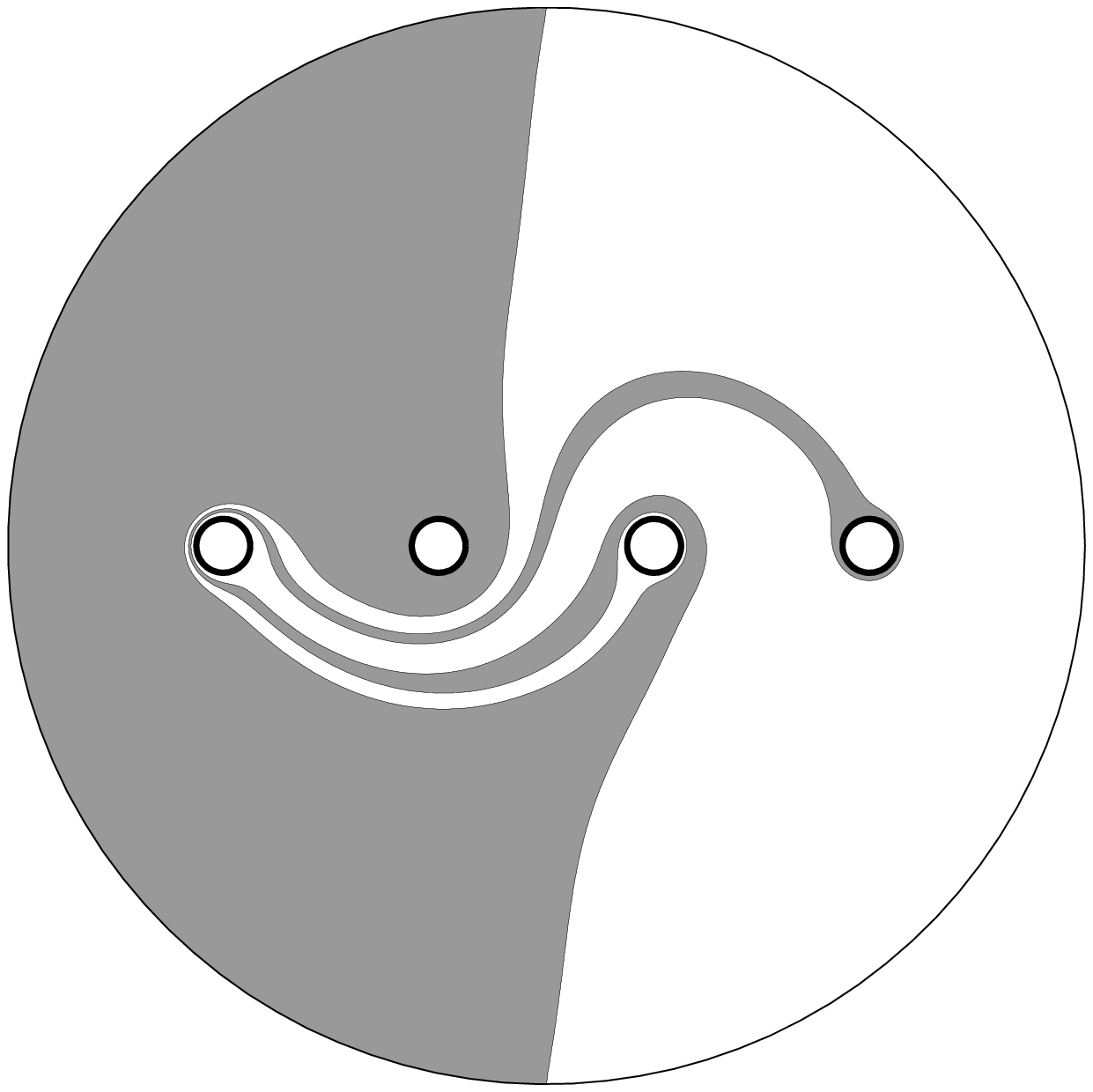}}
\parbox[t]{3.0cm}{\includegraphics[width=2.85cm]{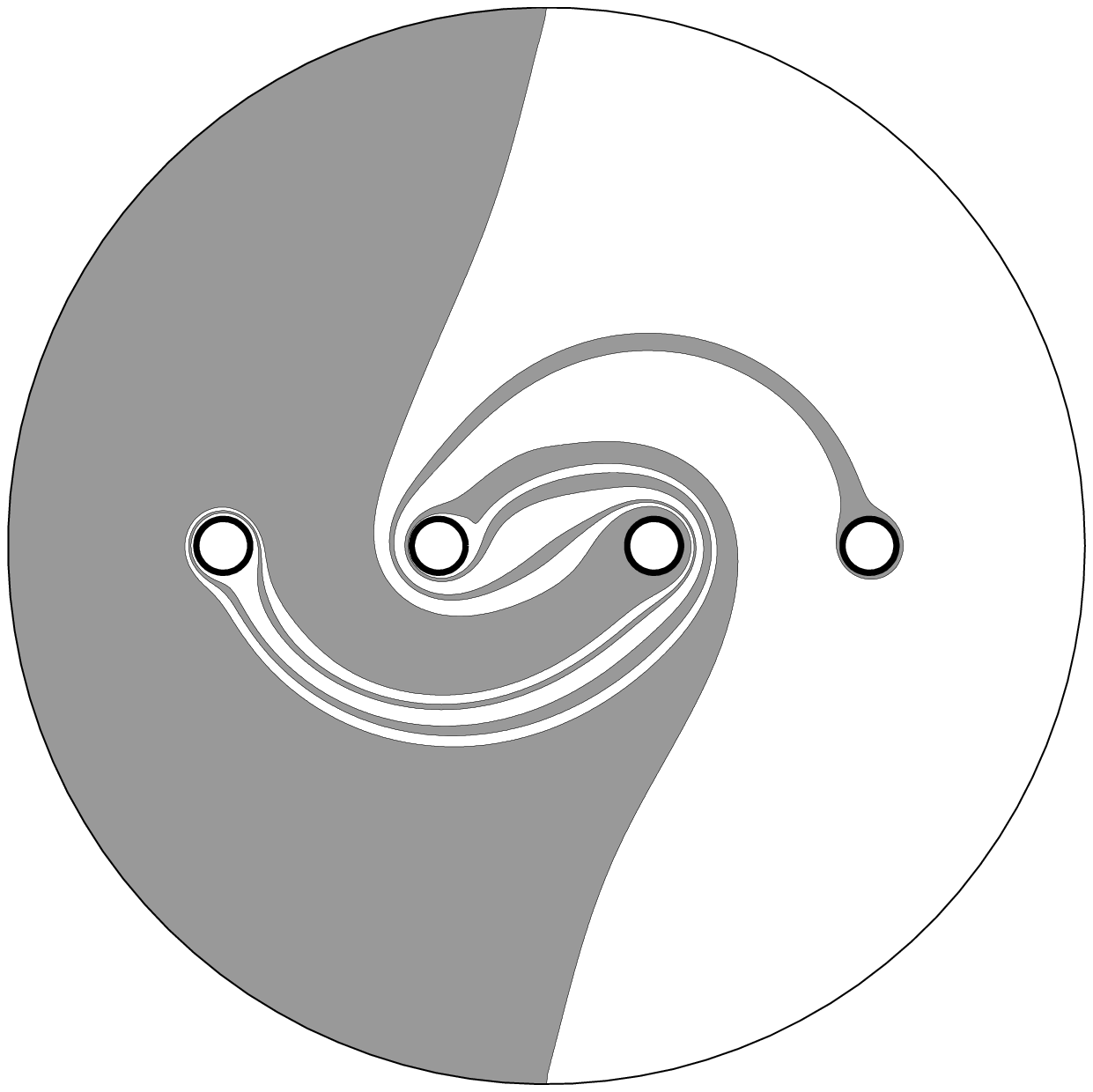}}
\parbox[t]{3.0cm}{\includegraphics[width=2.85cm]{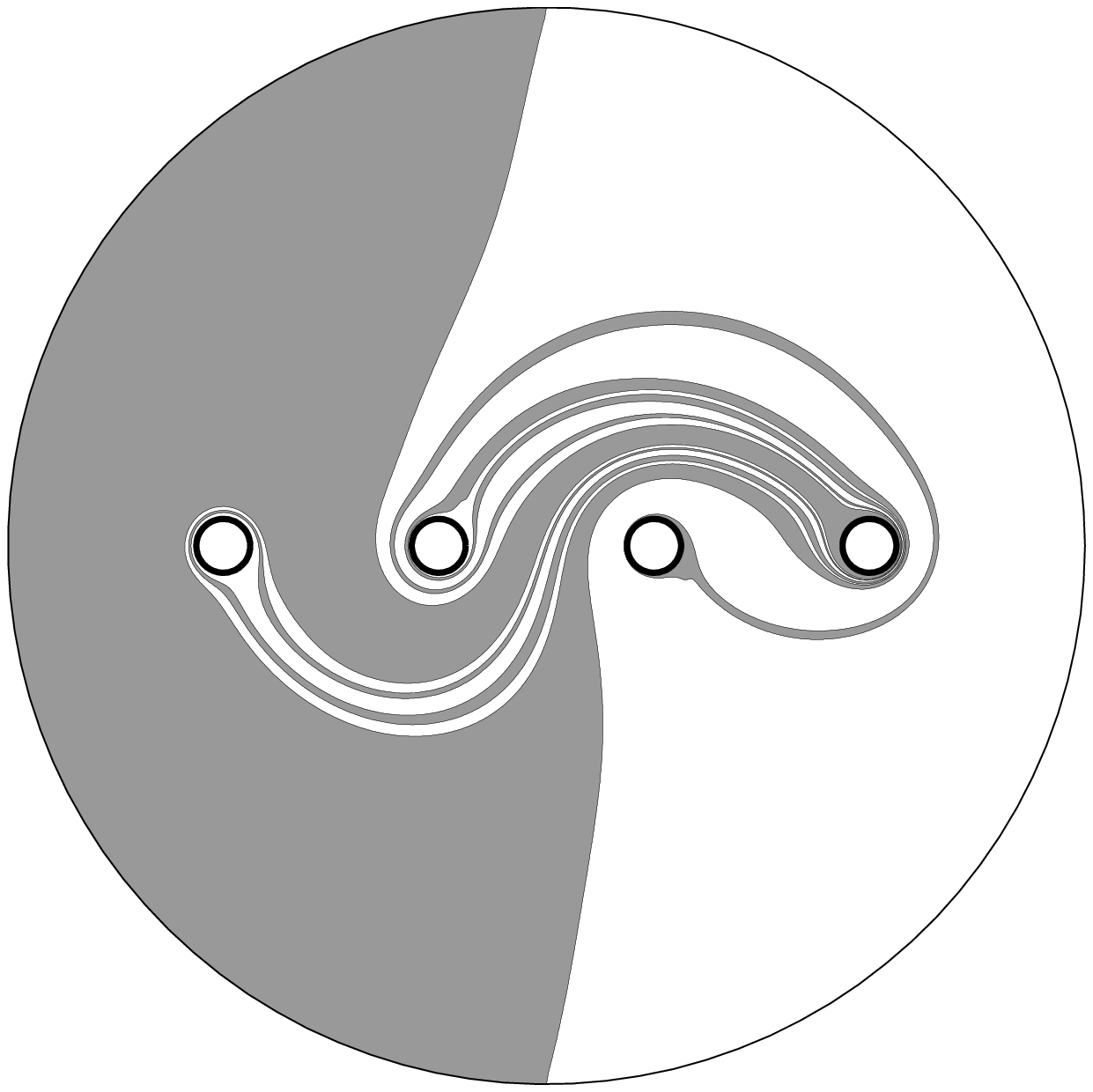}}
\parbox[t]{3.0cm}{\includegraphics[width=2.85cm]{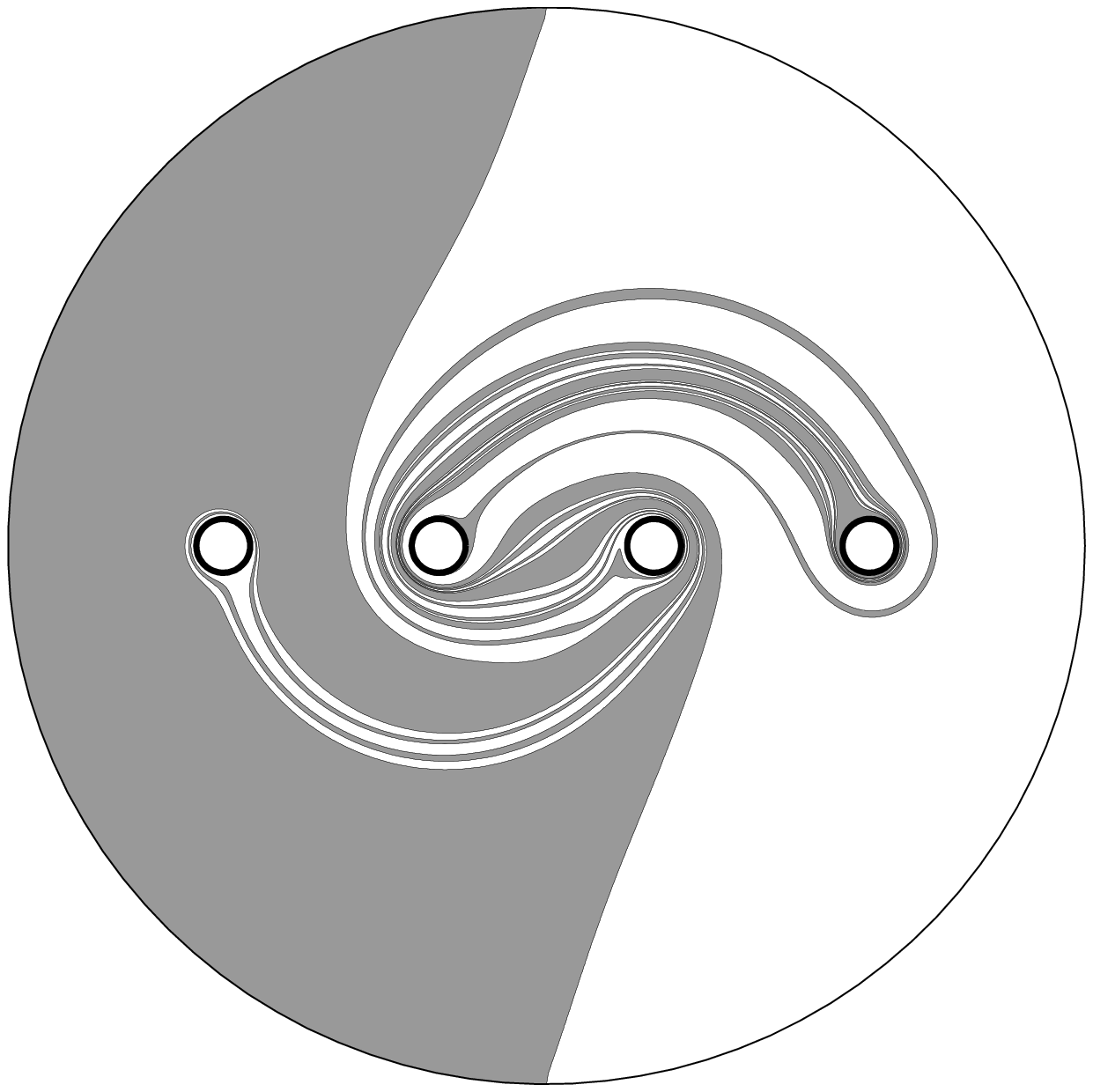}}
\end{center}
\caption{Stirring rods undergoing the braiding
  motion~$\sigma_1\sigma_{2}^{-1}\sigma_3\sigma_{2}^{-1}$ twice.  The
  generators are read from left to right, and each row is a full cycle of the
  braid.}
\label{fig:rods_s1s-2s3s-2}
\end{figure}

In \citet{Boyland2000} as well as subsequent
studies~\citep{MattFinn2003,Vikhansky2004} a three-rod mixer was considered,
as this is the minimum number of rods necessary to guarantee topological
chaos.  For variety, here we describe a four-rod device, since it has similar
properties but has not been described before.  In \S\ref{sec:opt} we will see
that this device is in some sense as `efficient' as the three-rod device.
Figure~\ref{fig:rods_s1s-2s3s-2} shows the effect of four stirring rods on a%
\begin{figure}
\begin{center}
\parbox[t]{3.0cm}{\includegraphics[width=2.85cm]{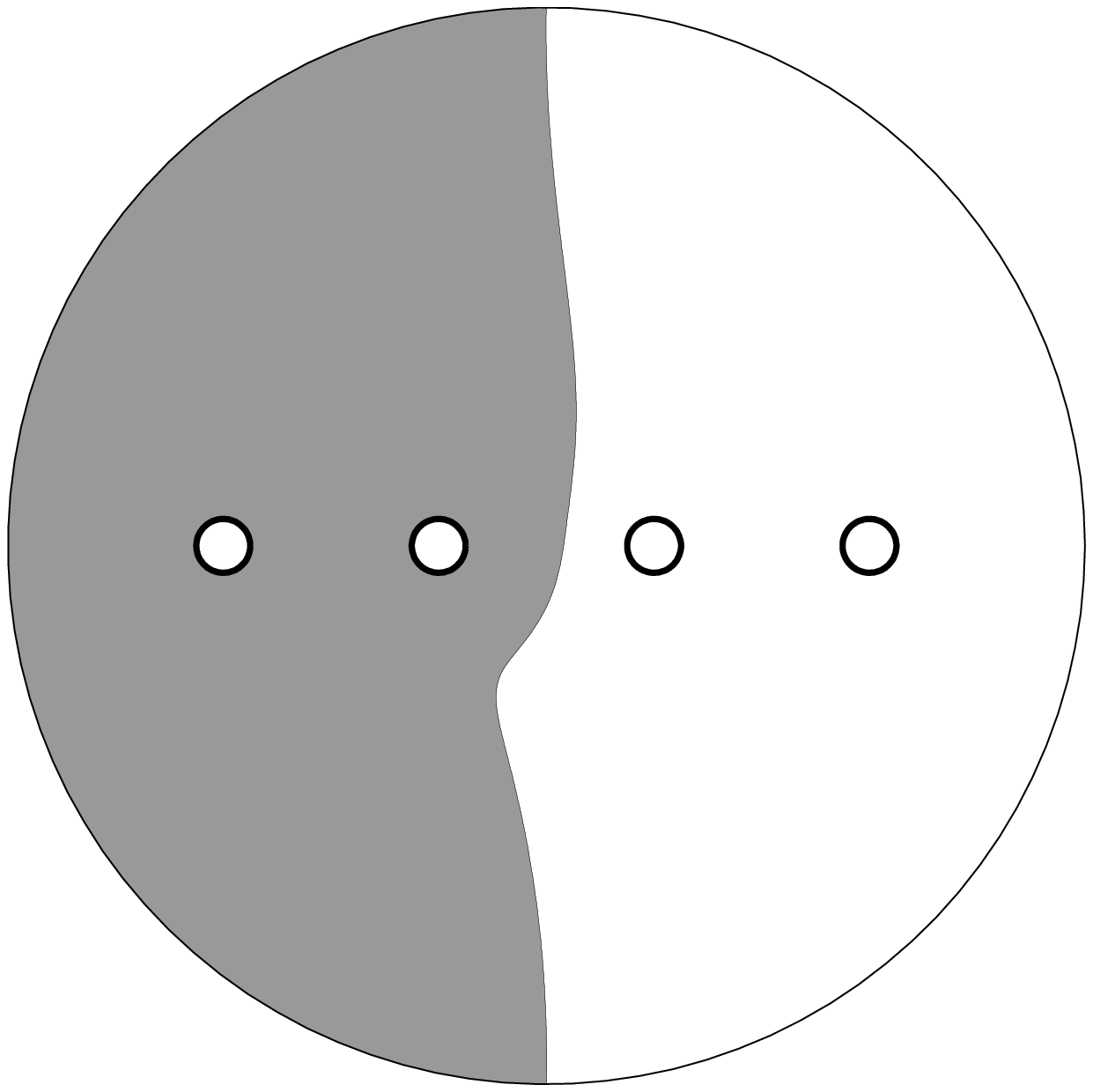}}
\parbox[t]{3.0cm}{\includegraphics[width=2.85cm]{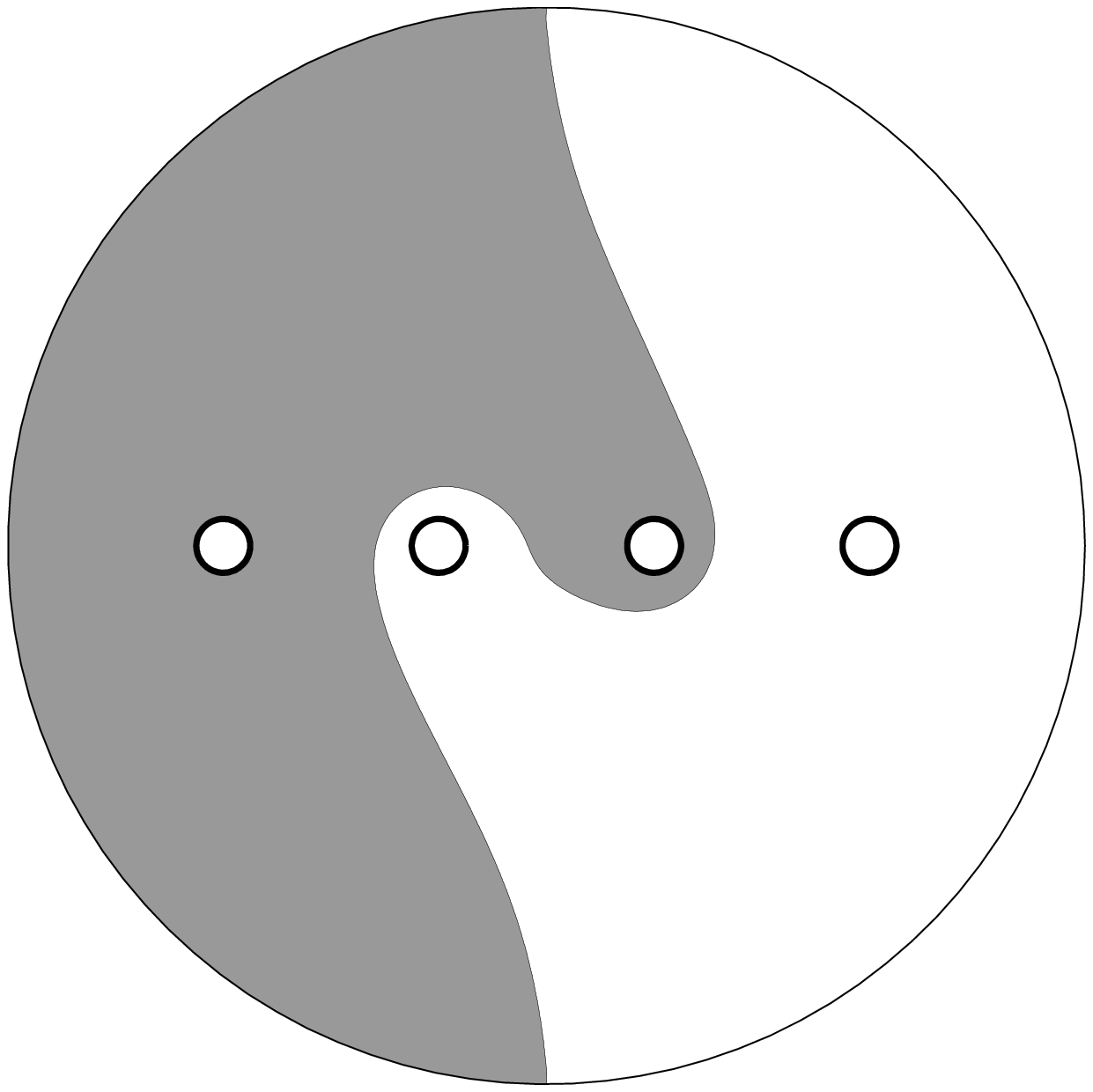}}
\parbox[t]{3.0cm}{\includegraphics[width=2.85cm]{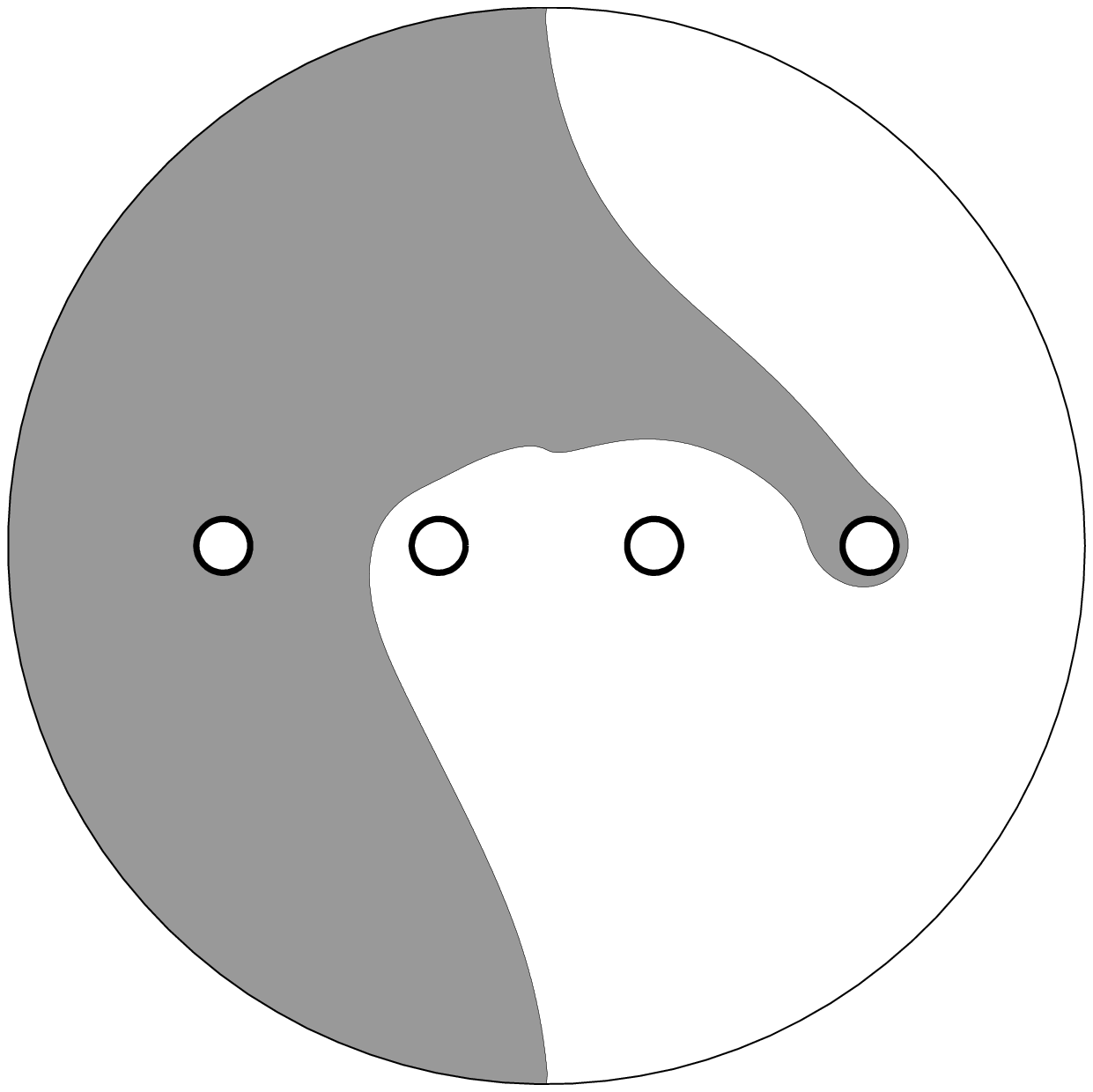}}
\parbox[t]{3.0cm}{\includegraphics[width=2.85cm]{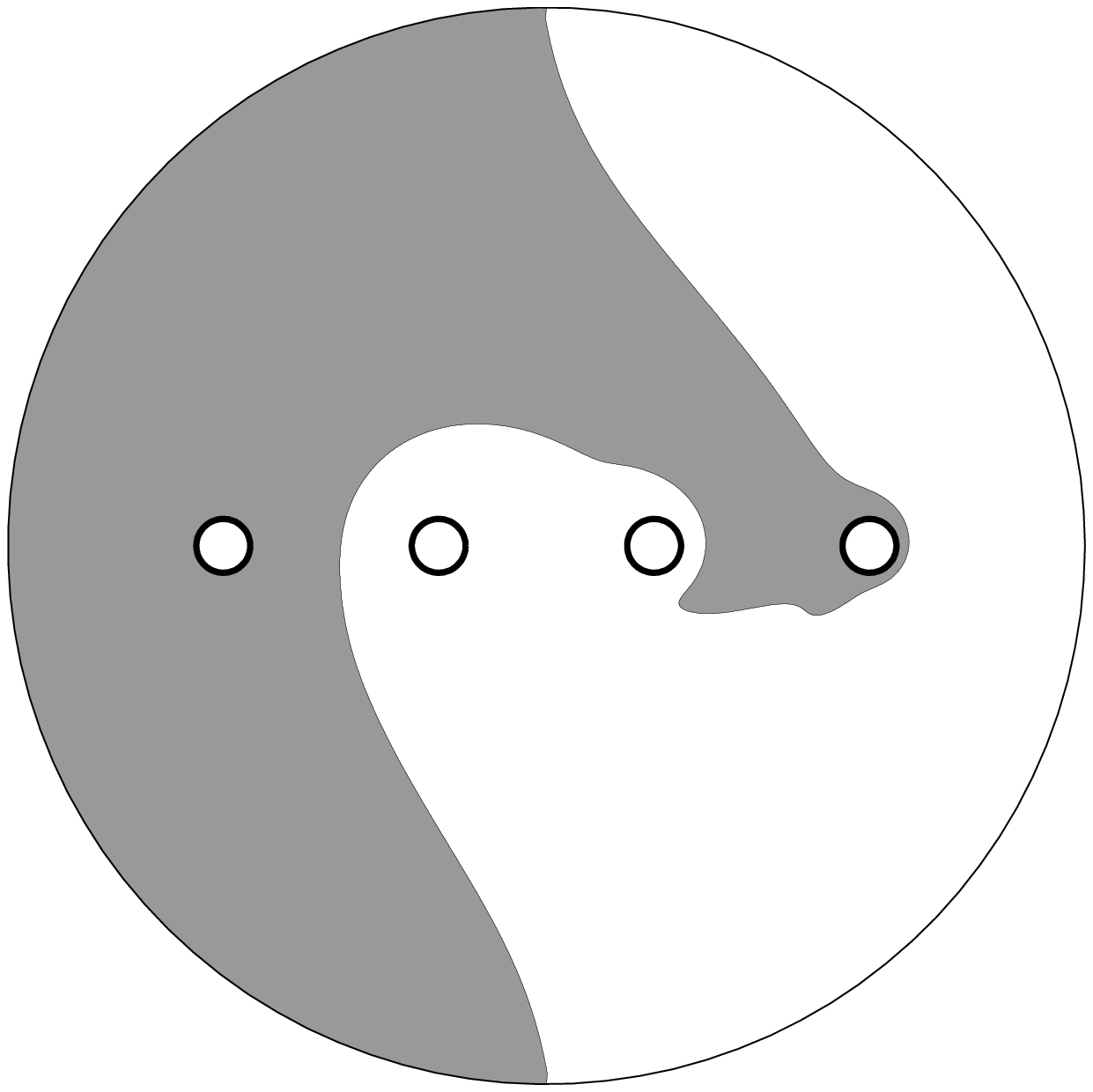}}

\smallskip

\parbox[t]{3.0cm}{\includegraphics[width=2.85cm]{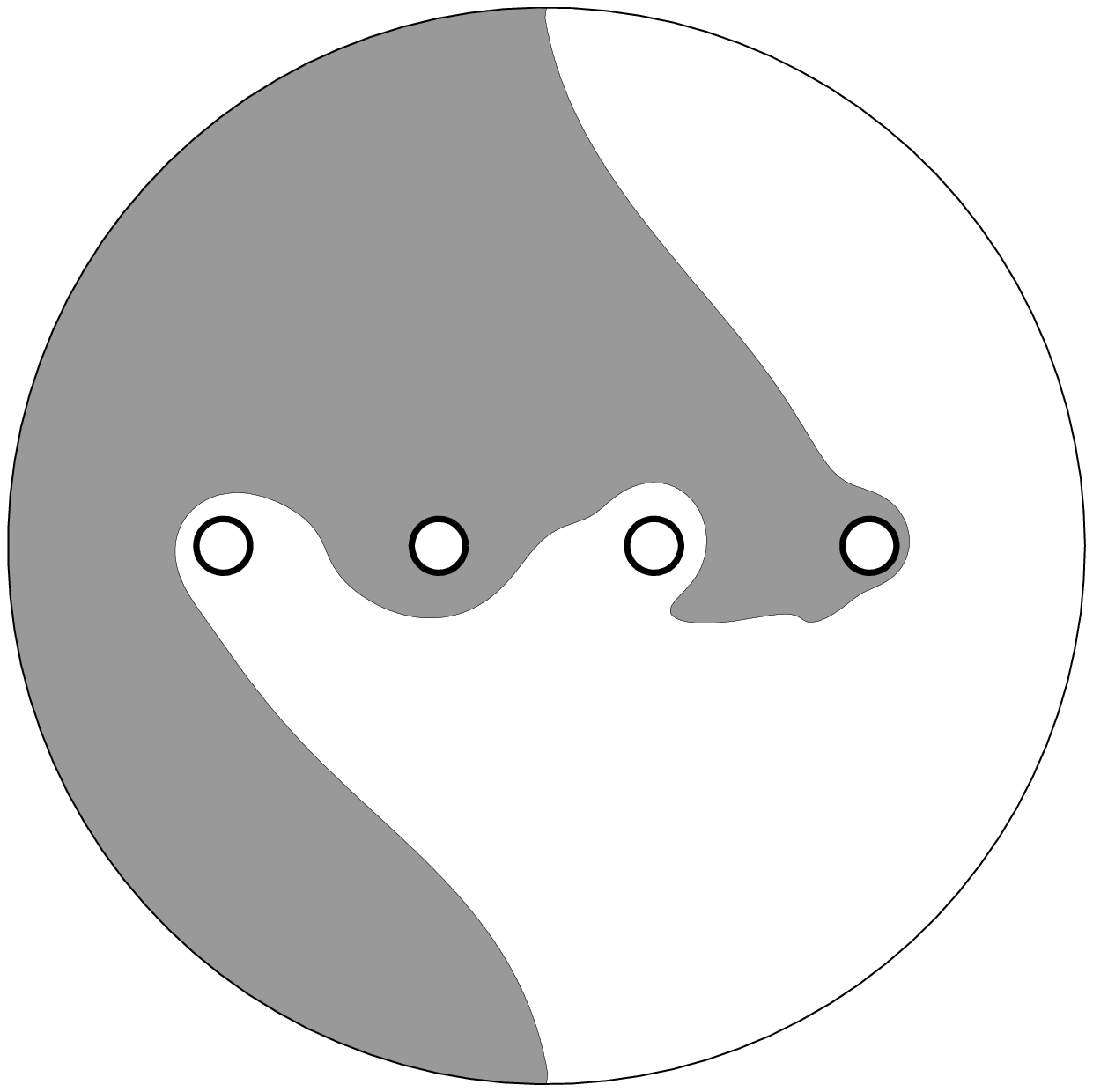}}
\parbox[t]{3.0cm}{\includegraphics[width=2.85cm]{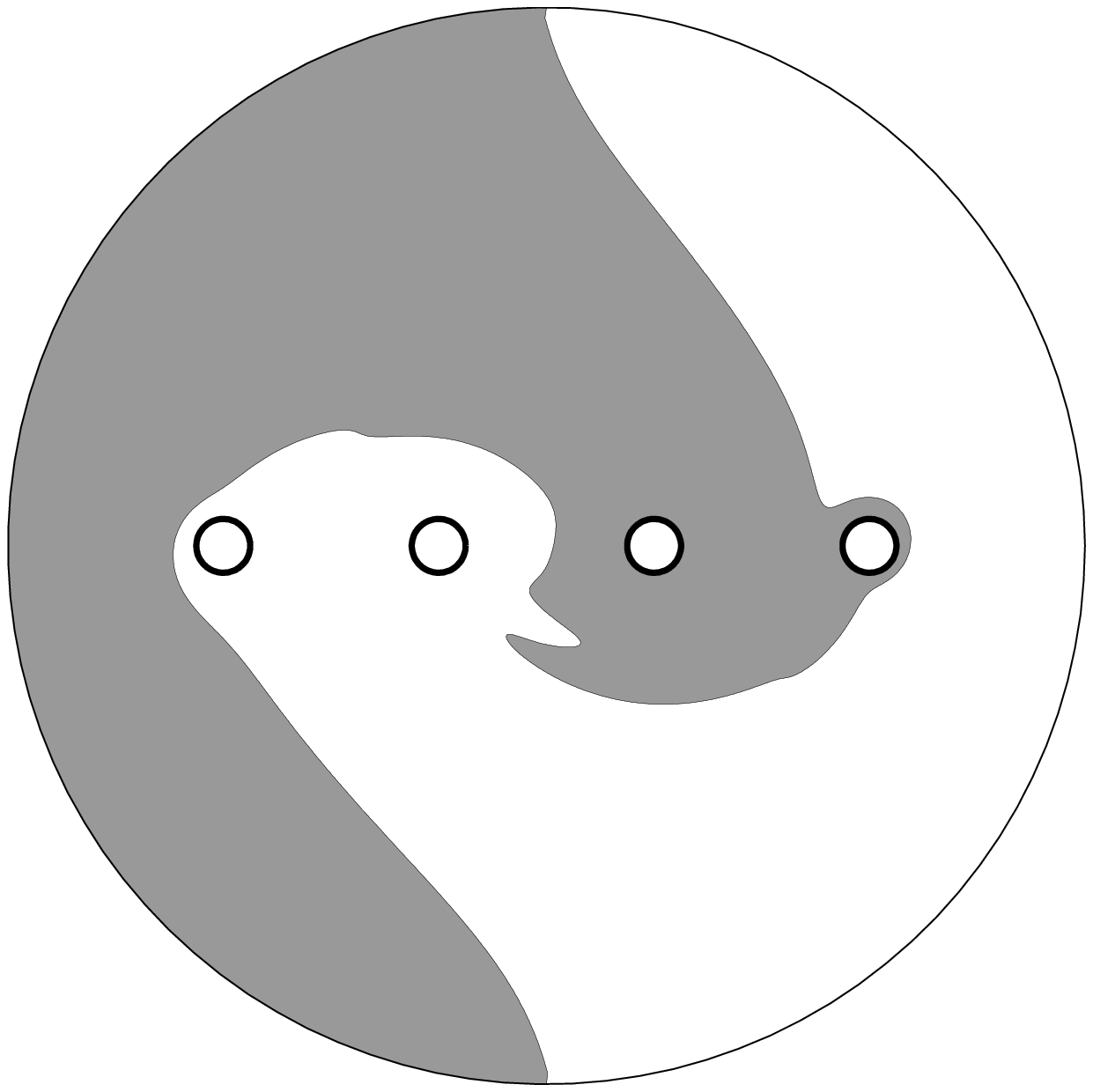}}
\parbox[t]{3.0cm}{\includegraphics[width=2.85cm]{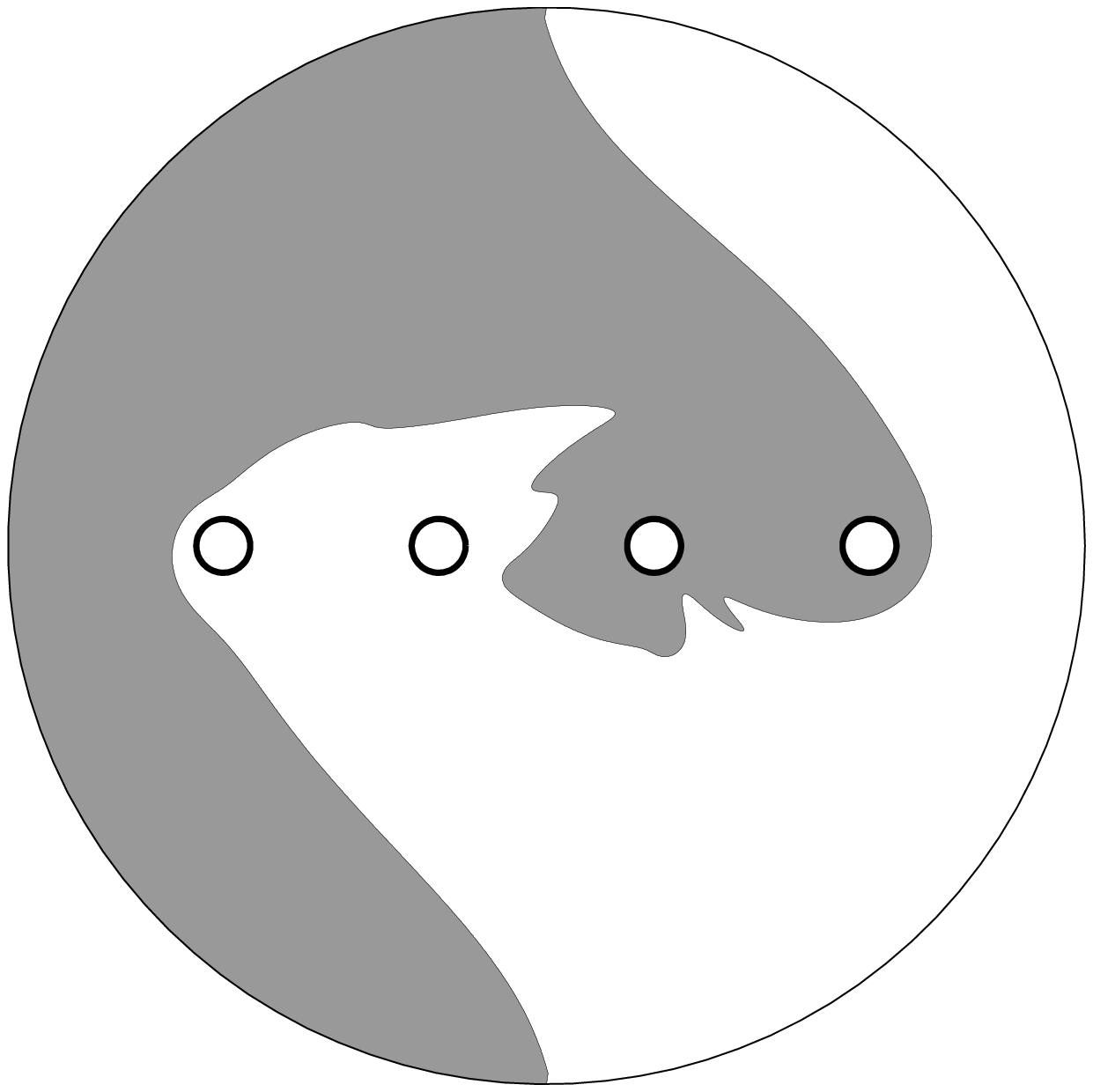}}
\parbox[t]{3.0cm}{\includegraphics[width=2.85cm]{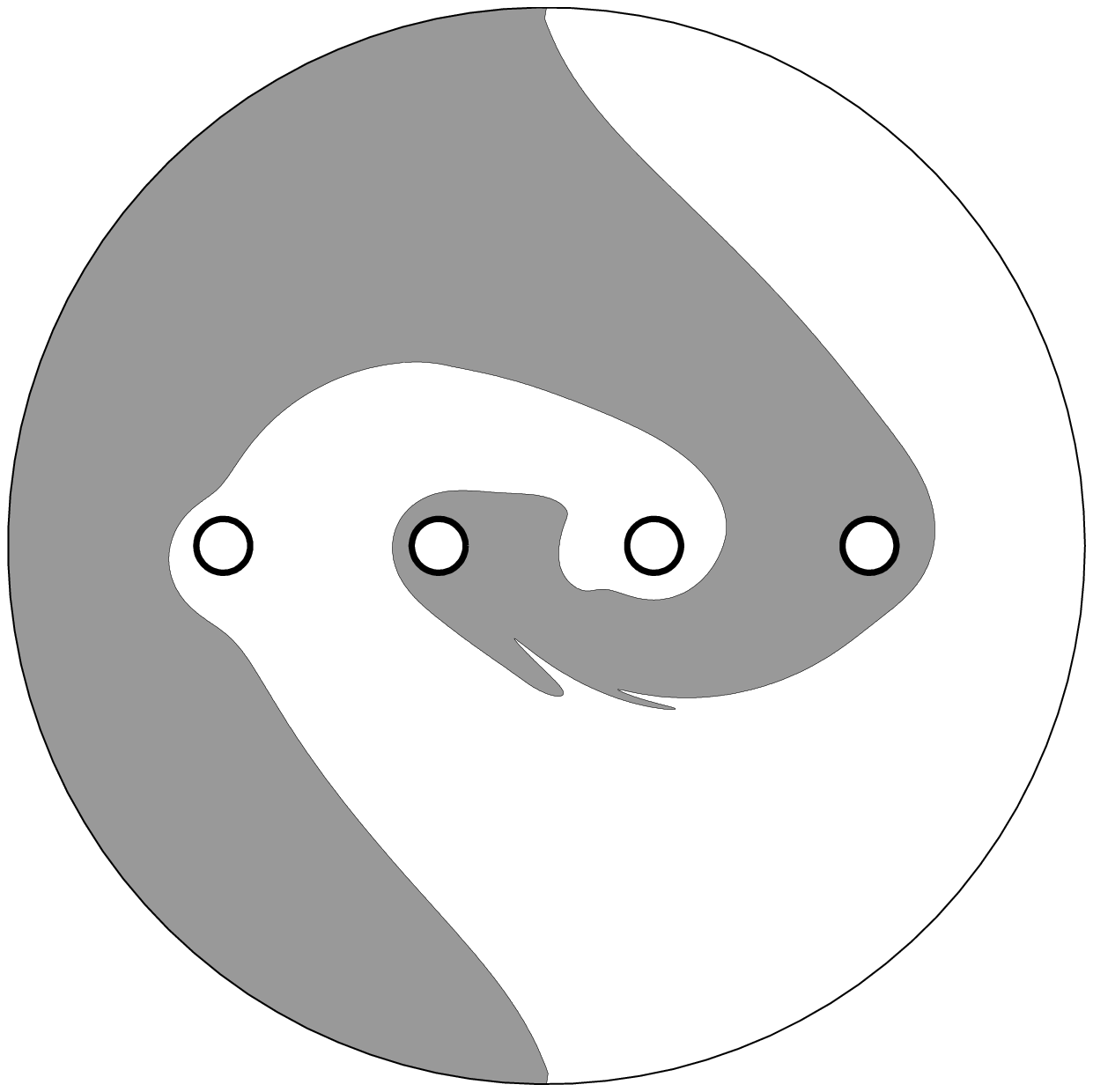}}
\end{center}
\caption{As in figure~\ref{fig:rods_s1s-2s3s-2}, but for the finite-order
  braid~$\sigma_1\sigma_{2}\sigma_3\sigma_{2}$.}
\label{fig:rods_s1s2s3s2}
\end{figure}
material line.  The rods are undergoing a motion specified by the
braid~$\sigma_1\sigma_{2}^{-1}\sigma_3\sigma_{2}^{-1}$, and the flow is
viscous Stokes flow.  This braid is pA (type 2 in
\S\ref{sec:topofluid}\ref{sec:TN}) with dilatation~$\lambda=(7+3\sqrt{5})/2
\simeq 6.8541$, which means that, asymptotically in time, a material line must
be stretched by at least this factor for every full period of the motion.
This is extremely efficient exponential stretching, and we will see in
\S\ref{sec:opt} that this particular~$\lambda$ is in a sense the optimal
efficiency.  Figure~\ref{fig:rods_s1s-2s3s-2} shows that the diffeomorphism
$\f$ contains a chaotic type~2 (pA) region, clearly visible as two crescents
in the centre, surrounding the rods.  It is fair to say that even though the
exponential stretching in the region is very large, the mixing region is
localised around the stirrers, at least for short times: topological
considerations predict nothing about the size of the pA region.  In
\S\ref{sec:opt} we will return to this issue.

In contrast, figure~\ref{fig:rods_s1s2s3s2} shows the action of stirring rods
moving according to the braid $\sigma_1\sigma_2\sigma_3\sigma_2$ on a material
line.  The length of the line is growing much more slowly, since this braid
specifies a finite-order isotopy class.  Note that the motions in
figure~\ref{fig:rods_s1s-2s3s-2} and~\ref{fig:rods_s1s2s3s2} have the same
energetic cost because of symmetry, but the former is far better at mixing.

This suggests that braids that specify pA isotopy classes imply better mixing
than finite-order ones.  All other things being equal, this is true, but we
also see that it is not sufficient: the size of the mixed region in
figure~\ref{fig:rods_s1s-2s3s-2} is fairly small.  In section~\ref{sec:opt} we
will focus on the issue of \emph{optimising} the braid for best mixing.  But
first we will see what can be learned from looking at periodic orbits of a
flow.

\section{Braids as a Diagnosis Tool: Ghost Rods}
\label{sec:diag}

So far we have looked at producing braids by moving rods in a two-dimensional
fluid.  The rods, being physical obstacles to material lines, induce a minimum
amount of stretching (topological entropy).  But any fluid particle orbit is a
topological obstacle to material lines.  What can we then deduce about the
mixing properties of a flow by studying the motion of fluid particles?  There
are two cases of interest, depending on whether we focus on periodic orbits or
chaotic orbits.

\subsection{Periodic Orbits}

A typical two-dimensional flow will contain \emph{chaotic regions} and
\emph{regular islands}.  The islands are said to be surrounded by a chaotic
sea, and for time-periodic flows they are best represented using a
stroboscopic map (Poincar\'e section), which is a snapshot of particle
positions at each period of the fluid motion.  However, a stroboscopic map
hides important information, especially that the islands are moving in between
snapshots.  In fact, since they form topological obstacles, the islands can be
regarded as playing the role of stirring rods.  Of course, they are not `true'
stirring rods, but we can deduce from their motion important information about
the nature of the chaotic sea around the islands.  In
figure~\ref{fig:tubeplot} we show the braid formed by regular islands and one
stirring rod in a viscous fluid.  The topological entropy of this braid
is~$70\%$ of the measured line-stretching exponent of the chaotic region.  (It
cannot exceed~$100\%$ since it is a lower bound on the line-streching
exponent.)  In that sense, these islands account for most of the chaos in the
region.  The rest is accounted for by including unstable periodic orbits in
the braid, but these are less useful since they are difficult to measure in
practical situations~\citep{Gouillart2006}.

\citet{Boyland2003} have also studied periodic braids formed by the motion of
point vortices.  Since vortices move as fluid particles, they are topological
obstacles to material lines.  They were thus able to show analytically that
chaos must occur.  Similarly, \citet{MattFinn2005preprint} have shown that
chaos must occur in the well-known sine flow by following periodic orbits on
the torus.

\subsection{Chaotic Orbits}
\label{sec:chaorb}

Another way to use braids as a diagnosis tool is to measure the braid formed
by arbitrary orbits~\citep{Thiffeault2005}, periodic, chaotic, or
otherwise. The great advantage of this approach is that it can be
realised experimentally by sprinkling tracer particles in a flow.  The
disadvantage is that the meaning of such `random' braids is not on a firm
mathematical footing, but the results suggest some interesting directions to
pursue (see also~\citet{Gambaudo1999,Boyland1994}).

The method is as follows: follow a large number of orbits in a flow, and as
they evolve compute the braid associated with their world lines.  The braid
word recording the crossings becomes longer and longer, but it never repeats
if the orbits are not periodic.  However, the topological entropy of such a
braid appears to \emph{converge} to a measurable value, in a similar manner to
Lyapunov exponents.  This topological entropy is a lower bound for the
line-stretching exponent of the flow, but the question is how sharp.  There
are encouraging results that using a large number of particles allows one to
measure the topological entropy of a flow fairly
accurately~\citep{MattFinn2006b}.  This has real practical advantages since the
topological entropy is difficult to measure by other means.

Until recently, another difficulty lay in computing the topological entropy
for very long braids (with thousands of generators) over many strands.  The
Burau representation~\citep{Kolev1989} offers an efficient method of computing
the entropy, but it only provides a lower bound for more than three strands
(and usually not a very good one, as we found in practice).  The
Bestvina--Handel algorithm mentioned in~\S\ref{sec:enforcing}\ref{sec:goodbad}
is accurate, but prohibitively expensive even for moderately long braids
(forty generators or so).  The best new algorithm so far is due to
\citet{Moussafir2006}: one records the crossings of a curve using
coordinates introduced by~\citet{Dynnikov2002}.  The number of crossings
yields a sequence that rapidly converges to the topological entropy.

\section{Braids as a Design Tool}
\label{sec:opt}

We have discussed in \S\ref{sec:enforcing}--\ref{sec:diag} how braids can be
used to enforce a minimum amount of stretching in a flow, and how they can be
used to diagnose its chaotic properties.  Now we turn to a more direct
application: can the braiding properties of a set of rods be optimised to give
the best possible lower bound on stretching?  We will see that, as in many
optimisation problems, part of the difficulty lies in formulating the question
properly.

\subsection{Optimising over Generators}
\label{sec:optgen}

The most basic optimisation problem is this:
\begin{quote}
  What word of length~$\nword$ maximises the entropy in the braid
  group~$\Br_\nn$?
\end{quote}
Clearly, as~$\nword$ gets larger, the maximum entropy also increases since we
have more generators to play with.  For~$\nn=3$, \citet{DAlessandro1999}
proved that the optimal entropy lies in repeating the
word~$\sigma_1\sigma_2^{-1}$ over and over again.  The dilatation of this
braid is~$(3+2\sqrt{5})/2$.  This is the square of the Golden
ratio~$(1+\sqrt{5})/2$, and we refer to the braid~$\sigma_1\sigma_2^{-1}$ as
the \emph{golden braid}.  Hence, the \emph{entropy per generator} of the
optimal braid in~$\Br_3$ is always equal to the logarithm of the Golden ratio.

What does this mean?  The optimal braid arises from a sequence where rods
rotate alternately clockwise and counterclockwise.  The resulting `pile-up' of
material lines can be related to the continued fraction expansion of the
Golden ratio,~$\{1,1,1,1,\ldots\}$.  We also note that the entropy per
generator associated with the protocol~$\sigma_1^m\sigma_2^{-m}$
is~$m^{-1}\log((m+\sqrt{m^2+4})/2)$.  The argument of the log gives the
so-called \emph{silver means}.%
\footnote{As for the Golden ratio, there is a geometrical construction of the
  silver means: start with a rectangle with one side of unit length, and
  remove~$m$ unit squares.  The ratio of the sides of the remaining rectangle
  is given by the~$m$th silver mean if it is the same ratio as the original
  rectangle.}
These are generalisations of the Golden ratio and have continued fraction
representation~$\{m,m,m,m,\ldots\}$.  As far as we know, this connection
between continued fraction expansions and stirring protocols has not been
thoroughly investigated.  A special case of the silver means, $m=2$, is often
called the \emph{silver ratio} and will be important in
\S\ref{sec:opt}\ref{sec:silvermix}.

Here are some further conjectures about optimal braids.  We discovered these
by using computer programs to examine large number of braids, but have no
rigorous proofs.  (Note that \citet{Moussafir2006} has independently
formulated these conjectures.)  For~$\nn=4$, the
braid~$\sigma_1\sigma_2^{-1}\sigma_3\sigma_2^{-1}$ has entropy per generator
equal to the Golden ratio.  (This is the `good' braid discussed in
\S\ref{sec:enforcing}\ref{sec:goodbad}.)  It is thus as effective
as~$\sigma_1\sigma_2^{-1}$ in~$\Br_3$.  For~$\nn>5$, \emph{all irreducible
braids have topological entropy per generator less than the logarithm of the
Golden ratio.}  So the Golden ratio braids only exist for~$\nn=3$ and~$4$, and
they give the highest possible entropy per generator.  Intuitively, these
braids involve a tight combination of three or four strands to achieve their
high entropy per generator.  More strands requires spurious switches down the
braid to make it irreducible.  A final unproved observation is that the braids
with the largest entropy always have alternating signs of noncommuting
generators, as in~$\sigma_1\sigma_2^{-1}\sigma_3\sigma_2^{-1}$.

Note that there is also an optimisation problem for irreducible braids with
\emph{minimum} (but positive) entropy.  Paradoxically, the minimum entropy
becomes smaller for larger~$\nn$, and it is easy to find sequences of braids
that have entropy decreasing as~$1/\nn$.  For example, the braid word
$\sigma_1\sigma_2\cdots\sigma_{\nn-2}\sigma_2\sigma_3\cdots\sigma_{\nn-1}$ has
entropy approximately equal to~$(2/\nn)\log(2+\sqrt{3})$, for~$\nn$ odd.
For~$\nn=5$, this braid has the least entropy of all irreducible
braids~\citep{Ham2006}.  The proof requires a computational train-track
automaton, and appears prohibitively difficult for higher~$\nn$.  Of course,
for our purposes here designing mixers based on the \emph{worst} braids is not
very useful, but it is a mathematically interesting question.

\subsection{Silver Mixers}
\label{sec:silvermix}

For practical applications, the type of optimisation discussed in
\S\ref{sec:opt}\ref{sec:optgen} is fundamentally flawed.  The main issue is
that the generators of the braid group,~$\sigma_i$, do not correspond to
natural motions of rods in physical systems.  For instance, the exchange of
the first and last rod in the four-rod system of
figure~\ref{fig:rods_s1s-2s3s-2} is
written~$\sigma_3\sigma_1\sigma_2\sigma_1\sigma_3$ in terms of exchanges of
neighbouring rods, so this simple operation requires five generators!  But
clearly it does not cost much more energy to do this than to exchange the
first and second rods ($\sigma_1$).  The generators do not capture the
intrinsic geometry of the system.

Another issue is that motions involving commuting generators (such
as~$\sigma_1$ and~$\sigma_3$ or any pair of nonadjacent generators) can be
performed simultaneously.  This is an advantage: energy is not usually the
most severe constraint in a real device, so contiguous commuting generators
should be weighted as one generator for the purposes of efficiency.  Many
different optimisation problems can thus be formulated to build in different
engineering constraints.  In particular, the resulting rod motions must be
realisable using a straightforward design.

\begin{figure}
\begin{center}
\includegraphics[width=.75\textwidth]{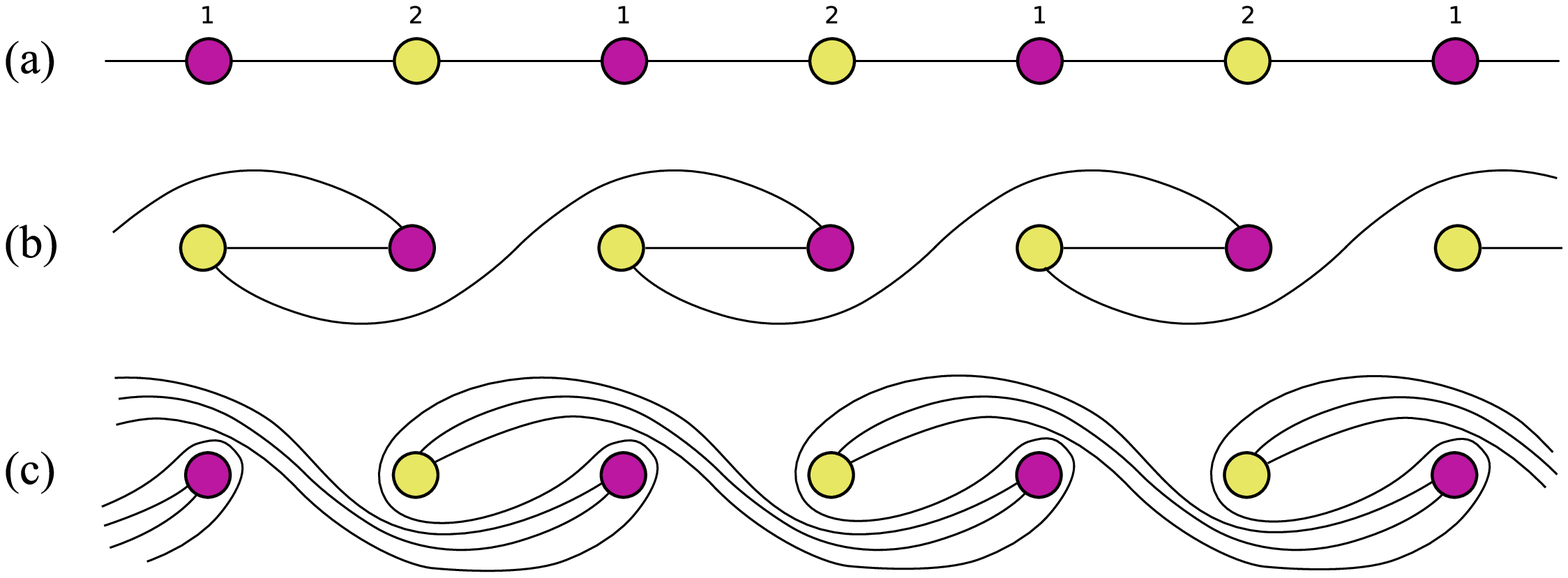}
\end{center}
\caption{(a) Two rods on a periodic lattice, joined by line segments. The
  segments are stretched by (b) a clockwise interchange of the rods at
  positions~$1$ and~$2$; and (c) a counterclockwise interchange of the rods at
  position~$2$ and~$1$.}
\label{fig:silver_lattice}
\end{figure}

Here we offer a more specialised approach that obviates these issues.
Consider a periodic lattice of two rods, as in
figure~\ref{fig:silver_lattice}(a).  Now perform a~$\sigma_1$ operation
(figure~\ref{fig:silver_lattice}(b)), followed by~$\sigma_2^{-1}$
(figure~\ref{fig:silver_lattice}(c)), where we redefine~$\sigma_2$ to mean a
clockwise interchange of rod~$2$ with the next rod to its right, which is the
periodic image of rod~$1$.  It is straightforward to compute the topological
entropy per generator of this braid,~$\log(1+\sqrt{2})$.  The
number~$1+\sqrt{2}\simeq 2.4142$ is the silver mean for~$m=2$ and is called
the silver ratio (it is considerably larger than the Golden ratio
$(1+\sqrt{5})/2\simeq 1.6180$).  For this reason, we refer to the
braid~$\sigma_1\sigma_2^{-1}$ for two rods in a periodic lattice as the
\emph{silver braid}.  That the silver braid has greater entropy per generator
than the golden braid does not contradict the optimality conjecture of
\S\ref{sec:opt}\ref{sec:optgen}, since that applied to a bounded domain,
whereas here we have a periodic array of rods.  We have recently examined
topological mixing in periodic and biperiodic geometries
\citep{MattFinn2005preprint,MattFinn2006b}.  A periodic array of two rods is
equivalent to arranging an even number of rods in a circle, with a fixed rod
in the centre to give the system the topology of an annulus.

\begin{figure}
\begin{center}
\parbox[t]{3.cm}{\includegraphics[width=3.0cm]{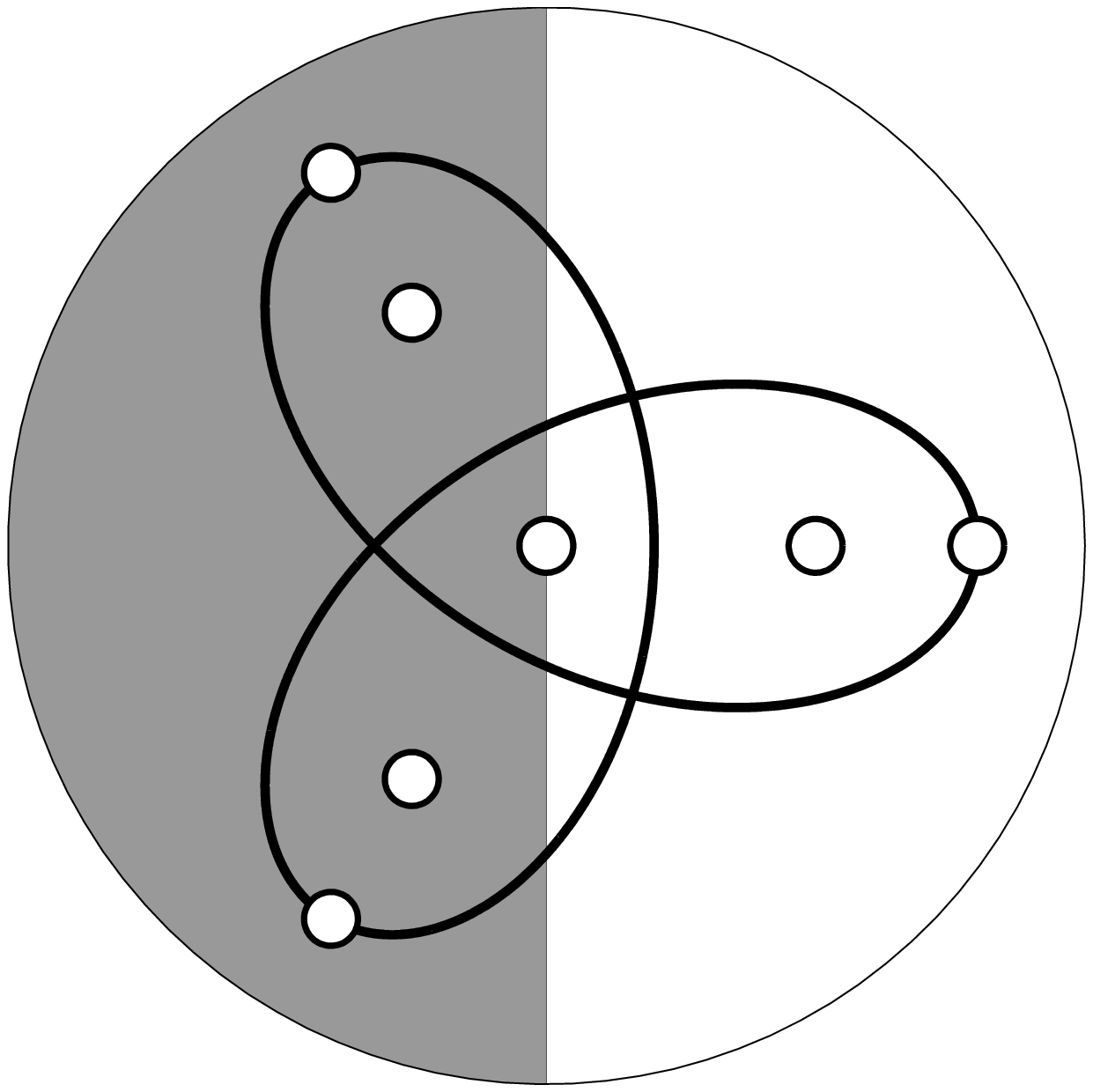}}
\parbox[t]{3.cm}{\includegraphics[width=3.0cm]{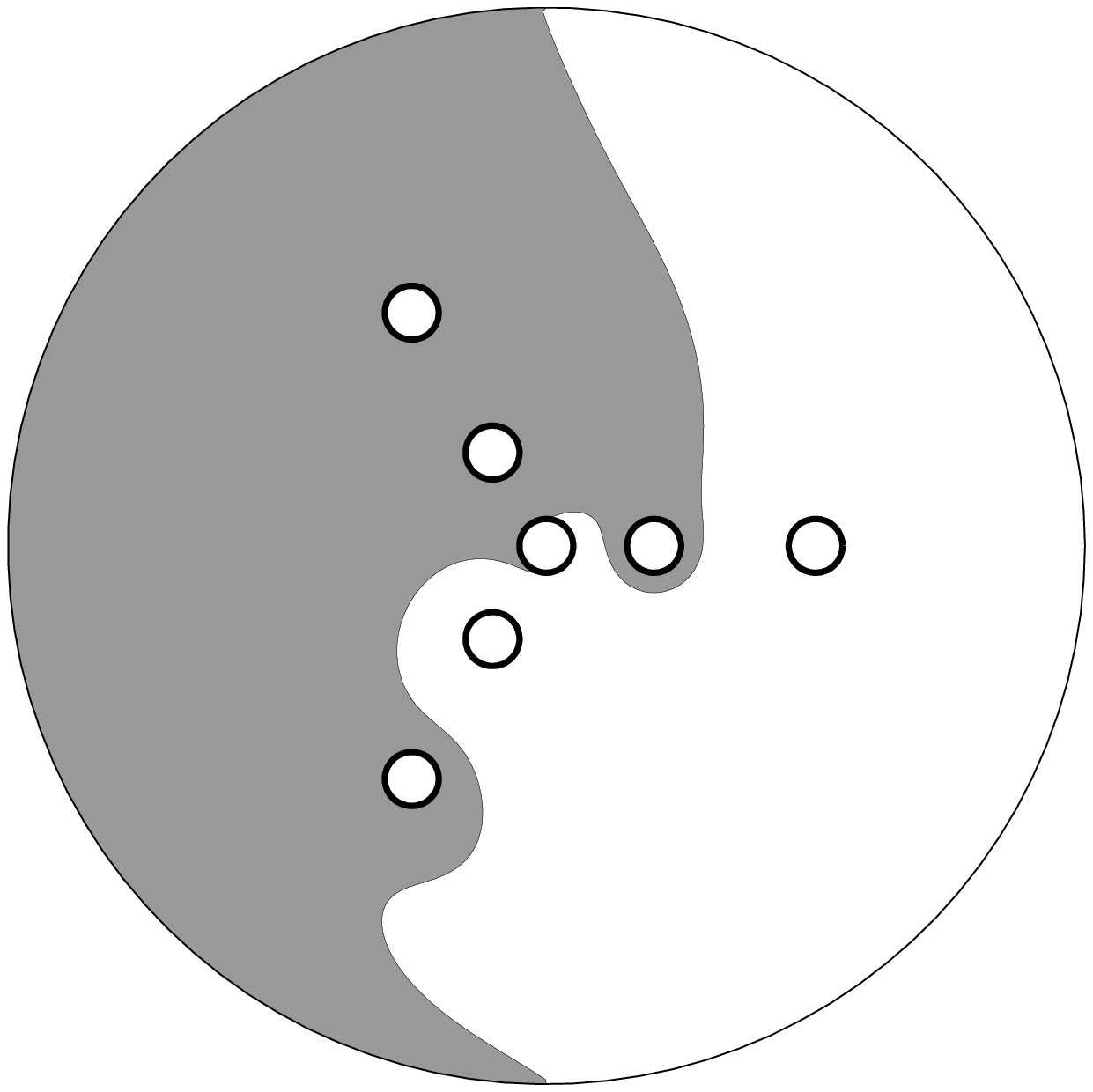}}
\parbox[t]{3.cm}{\includegraphics[width=3.0cm]{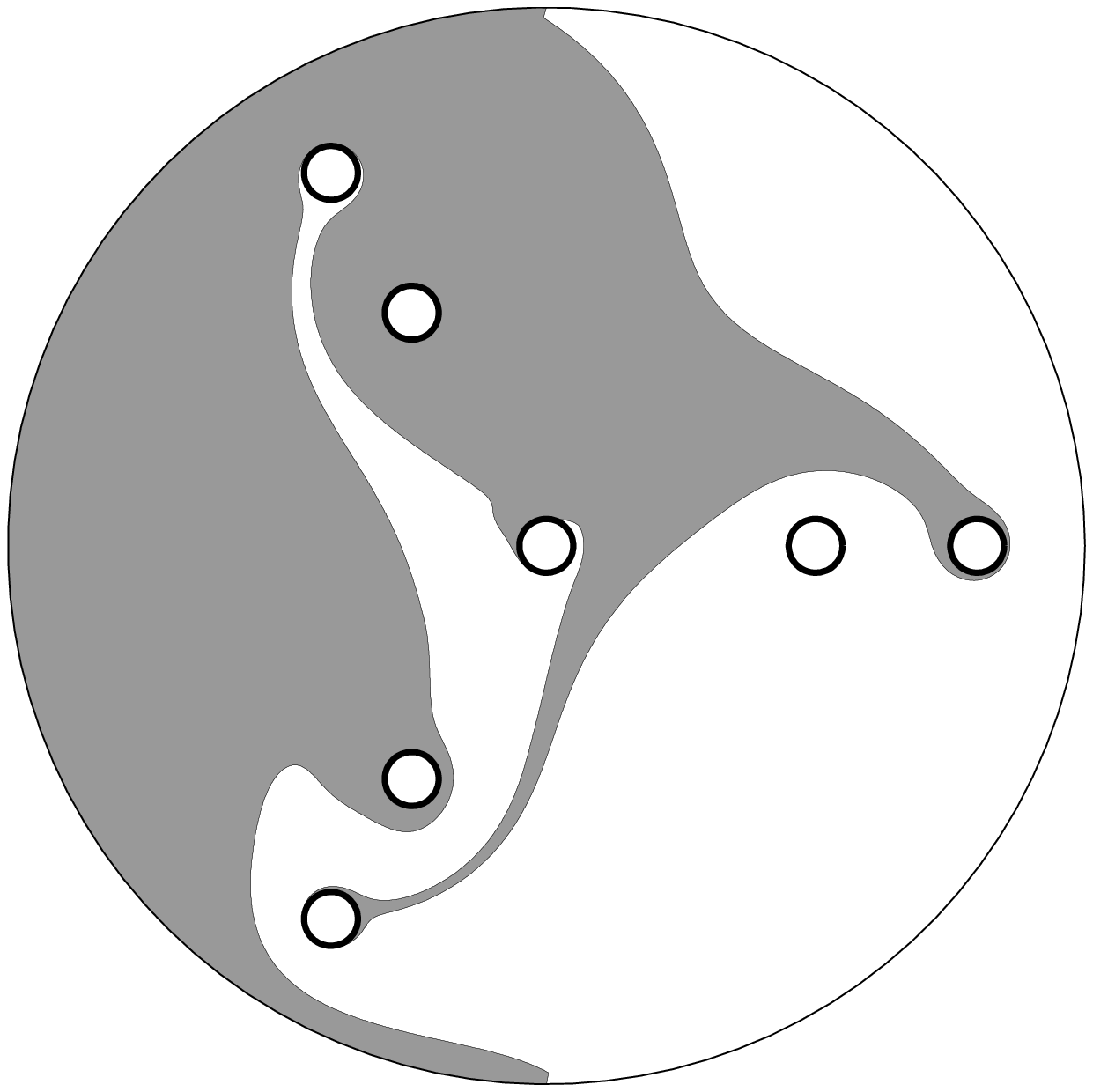}}
\parbox[t]{3.cm}{\includegraphics[width=3.0cm]{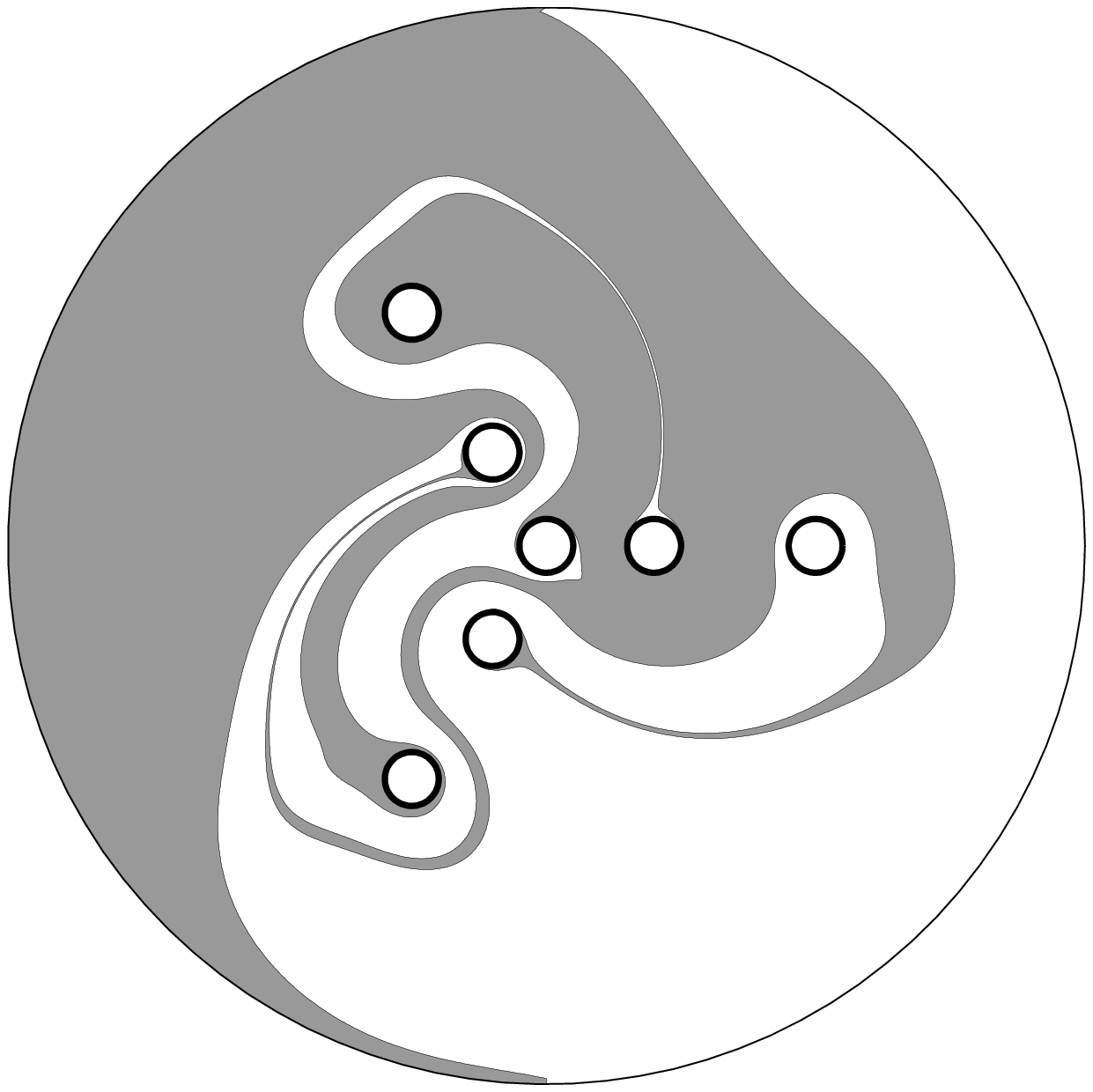}}

\smallskip

\parbox[t]{3.cm}{\includegraphics[width=3.0cm]{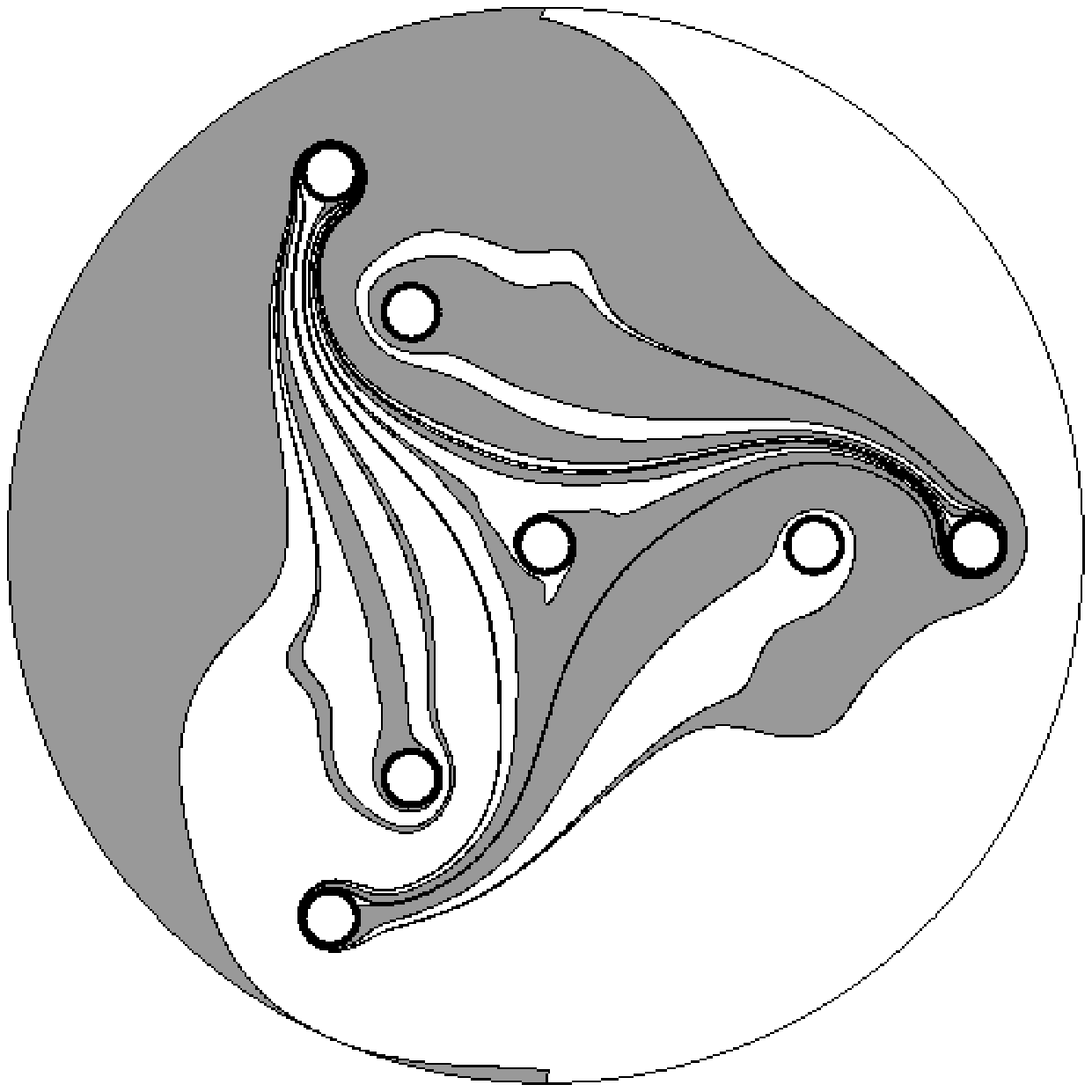}}
\parbox[t]{3.cm}{\includegraphics[width=3.0cm]{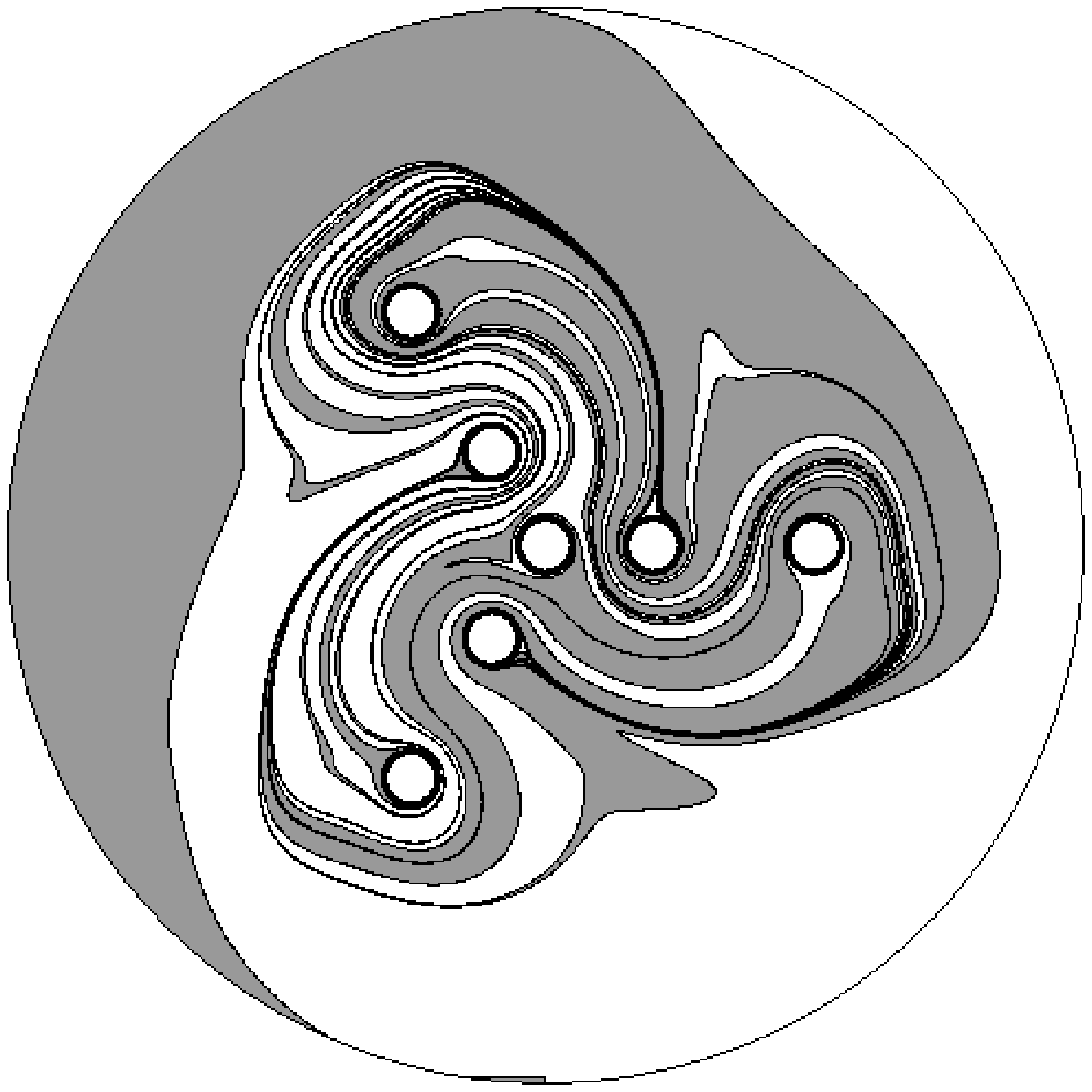}}
\parbox[t]{3.cm}{\includegraphics[width=3.0cm]{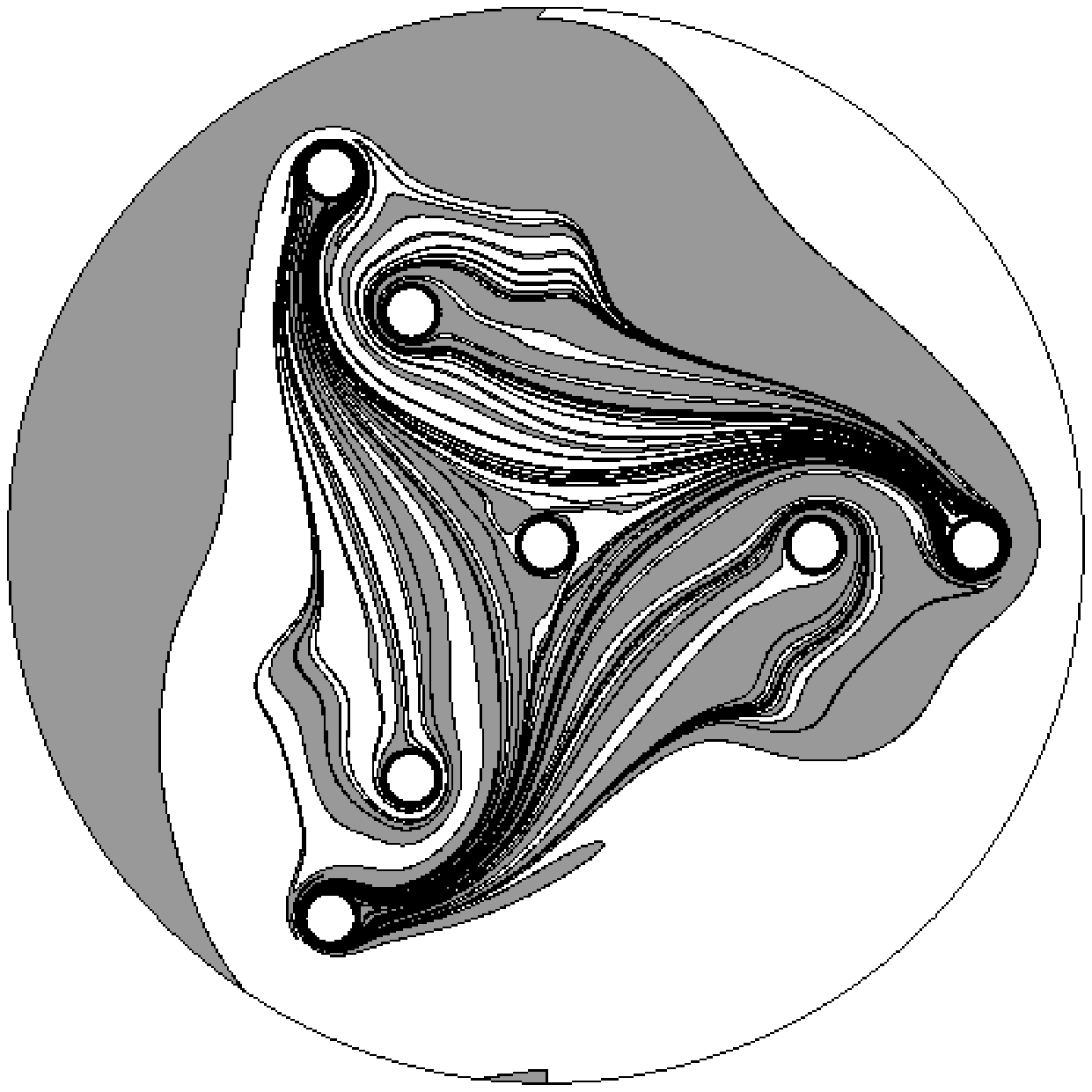}}
\end{center}
\caption{The silver mixing protocol: the central rod and three peripheral rods
are fixed, and the other three rods follow the same epicyclic path, shown in
the first snapshot.}
\label{fig:silvermix}
\end{figure}

The great advantage of this configuration is that it can be implemented
practically.  Figure~\ref{fig:silvermix} shows rod motions that are
topologically equivalent%
to the silver braid protocol.  The central rod defines the annular domain, and
three rods are fixed.  The remaining three rods move in an epicyclic motion as
shown in the first snapshot.  The net effect is that each moving rod executes
an `over-under' sequence with the fixed rods as it travels around the mixer,
which is exactly the silver braid.  As can be seen in the final snapshot,
there is a rather large central mixing region where material lines are
stretched exponentially fast, at a rate bounded from below
by~$6\log(1+\sqrt{2})$ for each full cycle, since the rods undergo
three~$\sigma_1\sigma_2^{-1}$ exchanges before returning to their initial
position.  More rods would lead to a greater topological entropy, but would
also complicate the apparatus.  Of course, topological entropy is not the only
important factor: a candidate protocol must also have a reasonably large
mixing region, and this can only be deduced by solving fluid equations
(viscous Stokes flow here).  Figure~\ref{fig:silvermix} shows that the silver
mixer has a large triangular mixing region, compared for instance with the
protocol in figure~\ref{fig:rods_s1s-2s3s-2}, which has a smaller S-shaped
mixing region.

\begin{figure}
\begin{center}
\includegraphics[width=.9\textwidth]{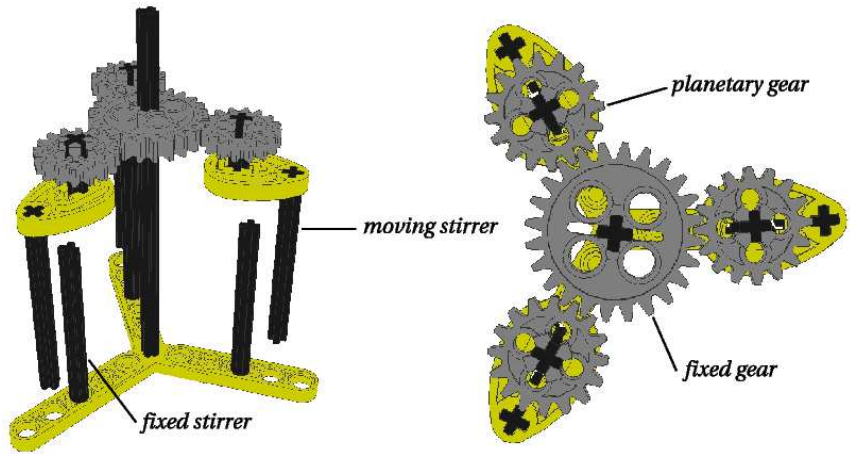}
\end{center}
\caption{An implementation of the silver mixing protocol in
  figure~\ref{fig:silvermix} using Lego\texttrademark, drawn with LeoCAD.  On
  the right is a detailed view of the gears.}
\label{fig:lego_mixer}
\end{figure}

The rod motion in the silver mixer may appear to be quite complicated, and
thus difficult to realise, but in figure~\ref{fig:lego_mixer} we show an
implementation using Lego\texttrademark\ building blocks.
The central rod supports the whole mechanism, and the three fixed rods are
attached to the base.  The epicyclic motion of the remaining three rods is
realised by having the correct gear ratio (16:24 here). The silver mixer is
thus practical from an engineering perspective.  By using the result of
\citet{DAlessandro1999}, the silver ratio entropy can be shown to be optimal
(per generator) for rods arranged in an annulus.

\section{The Future}
\label{sec:future}

We discuss a few promising areas for future research.

\emph{Three-dimensional flows.} The most important area of expansion for the
type of topological approach presented here is to develop a theory for
three-dimensional flows.  Knots and braids have been used in three-dimensional
flows for a long time, see for instance the book by~\citet{Ghrist}, usually
for steady flows. However, three-dimensional applications to mixing analogous
to \citet{Boyland2000} are not forthcoming.  To some extent, the theory is
already three-dimensional, in the sense that if rods extend all the way to the
bottom of the mixing device then the braid formed by the rods still gives a
bound on the topological entropy.  However, this cannot be applied to general
orbits.  The problem is that lifting three-dimensional particle orbits to a
four-dimensional space-time leads to trivial braids.  To put it simply, points
are not obstacles to lines in three dimensions.  Points are obstacles to
sheets, and one could imagine trying to record the motion of points as they
drag along two-dimensional sheets, but this does not put a lower bound on the
\emph{area} of the sheets.  The only option is to follow material lines
anchored to the boundary or loops, but so far this has not been very practical
and has not been developed.

It was speculated in \citet{Boyland2000} that rods could be inserted in pipe
mixers to take advantage of topological effects: the rods would mimic a braid
and force the fluid travelling down the pipe to have positive entropy.
\citet{MattFinn2003b} found this to be true, but the chaotic region was tiny
and confined near the rods, because of the no-slip boundary condition at rigid
surfaces.

\emph{Dynamics.} Another important area of future exploration is the
connection between effective braids and dynamics.  We need to understand
better the type of situations (geometry, etc.) that yield favorable braiding
motions of islands and periodic orbits, as well as produce large, uniform
mixing regions.  So far the best we can offer are case-by-case analyses, as in
\citet{Gouillart2006,Thiffeault2005preprint}, but this is not very
satisfactory.  In related work, \citet{Boyland2005} has recently investigated
the conditions under which Euler's equation produces time-periodic solutions
with pA dynamics.

\emph{Open Flows.} Most industrial situations involve open flows: fluid enters
a mixing region only for a finite time, and then exits, having hopefully
been mixed.  Can topological considerations tell us anything?  The
Thurston--Nielsen theorem does not apply, but we can define a topological
entropy by looking at the growth rate of material lines or the density of
periodic orbits.  The braiding motion of unstable periodic orbits (which make
up the chaotic saddle) can then be used to put lower bounds on the topological
entropy.

\begin{acknowledgements}

We thank Emmanuelle Gouillart and Jacques-Olivier Moussafir for their insights
into topological mixing.  This work was funded by the UK Engineering and
Physical Sciences Research Council grant GR/S72931/01.

\end{acknowledgements}

\setlength{\bibsep}{1pt}


\end{document}